\documentclass[manuscript,acmsmall,screen,nonacm]{acmart}

\AtBeginDocument{%
  }

\newcommand\blfootnote[1]{%
  \begingroup
  \renewcommand\thefootnote{}%
  \footnote{#1}%
  \addtocounter{footnote}{-1}%
  \endgroup
}

\usepackage{xurl}
\usepackage{amsmath,amssymb,amsfonts}
\usepackage{booktabs}
\usepackage{algorithm2e}
\usepackage{graphicx}
\usepackage{textcomp}
\usepackage{xcolor}
\usepackage{flushend}
\usepackage{microtype}
\usepackage{rotating}
\usepackage{tikz}
\usepackage{makecell}
\usepackage{enumitem}
\usepackage{wasysym}

\usepackage{multirow}
\usepackage{siunitx}

\definecolor{salmon}{HTML}{E50000}
\definecolor{lime}{HTML}{00E000}

\newcommand{\win}[2]{\ensuremath{(W,O)=(#1\,,#2)}}
\newcommand{\won}[2]{\ensuremath{(#1\,,#2)}}

\newcommand*\circled[1]{\tikz[baseline=(char.base)]{
            \node[shape=circle,draw,inner sep=0.7pt] (char) {#1};}}

\RestyleAlgo{ruled}
\SetKwComment{Comment}{/* }{ */}
\SetKwComment{Comment}{\textcolor{gray}{//} }{}
\SetKw{KwInitialize}{Initialize}
\SetKw{KwParse}{Parse}
\SetKw{KwRead}{Read Disk}
\SetKw{KwWrite}{Write Disk}
\SetKw{KwLoad}{Load}
\SetKw{KwRun}{Run}
\SetKw{KwLog}{Log}
\SetKw{KwUpdate}{Update}
\SetKw{KwResult}{Result}

\begin{document}

\title{{SHIELD}: A Host-Independent Framework for Ransomware Detection using Deep Filesystem Features}

\author{Md Raz}
\email{md.raz@nyu.edu}

\author{Venkata Sai Charan Putrevu}
\email{v.putrevu@nyu.edu}

\author{Prasanth Krishnamurthy}
\email{prashanth.krishnamurthy@nyu.edu}

\author{Farshad Khorrami}
\email{khorrami@nyu.edu}

\author{Ramesh Karri}
\email{rkarri@nyu.edu}
\affiliation{%
  \institution{Department of Electrical and Computer Engineering, Tandon School of Engineering, New York University}
  \city{Brooklyn}
  \state{New York}
  \country{USA}
  }

\blfootnote{This work is supported in part by the Defense Advanced Research Projects Agency (DARPA) under contract HR00112390029 and Department of Energy (DOE) under grant DE-CR0000051.}

\renewcommand{\shortauthors}{Raz et al.}

\begin{abstract}
Ransomware's escalating sophistication necessitates tamper-resistant, off-host detection solutions that capture deep disk activity beyond the reach of a compromised operating system.
Existing detection systems use host/kernel signals or rely on coarse block-I/O statistics, which are easy to evade and miss filesystem semantics. The filesystem layer itself remains underexplored as a source of robust indicators for storage-controller-level defense.
To address this, we present SHIELD: a 
    \underline{\textbf{S}}ecure 
    \underline{\textbf{H}}ost-\underline{\textbf{I}}ndependent \underline{\textbf{E}}xtensible Metric \underline{\textbf{L}}ogging Framework for Tamper-Proof \underline{\textbf{D}}etection and Real-Time Mitigation of Ransomware Threats. 
SHIELD parses and logs filesystem-level features that cannot be evaded or obfuscated to expose deep disk activity for real-time ML-based detection and mitigation.
We evaluate the efficacy of these metrics through experiments with both binary (benign vs. malicious behavior) and multiclass (ransomware strain identification) classifiers.
In evaluations across diverse ransomware families, the best binary classifier achieves 97.29\% accuracy in identifying malicious disk behavior. A hardware-only feature set that excludes all transport-layer metrics retains 95.97\% accuracy, confirming feasibility for FPGA/ASIC deployment within the storage controller datapath.
In a proof-of-concept closed-loop deployment, SHIELD halts disk operations within tens of disk actions, limiting targeted files affected to $\leq$0.4\% for zero-shot strains at small action-windows, while maintaining low false-positive rates ($\leq$3.6\%) on unseen benign applications. 
Results demonstrate that filesystem-aware, off-host telemetry enables accurate, resilient ransomware detection, including intermittent/partial encryption, and is practical for embedded integration in storage controllers or alongside other defense mechanisms.
\end{abstract}



\maketitle

\section{Introduction} 

    Ransomware, which renders critical cyber-physical infrastructure inoperative through data encryption and exfiltration, has evolved to leverage hardware-accelerated encryption, multi-threaded execution, and Ransomware-as-a-Service (RaaS) platforms to inflict massive financial and operational damage~\cite{cryptoviral, meland2020ransomware, CybotsAI2024}. 
    In 2024, over 5{,}461 successful attacks were reported, exceeding \$133 million in ransom payments with over \$40~billion in projected losses across healthcare, finance, and critical infrastructure sectors~\cite{ransomware_2024}. Reports show that frequency of ransomware attacks rose by 32\% in 2025, with 7{,}419 reported attacks worldwide, with an average ransom payment of about one million dollars per attack~\cite{cyberreport2025}.
    As many incidents remain undisclosed, these figures likely underestimate the true scope of the problem. 
    High-profile families like LockBit, RansomHub, BlackCat, and Rhysida dominate the threat landscape; LockBit reportedly encrypts up to 20,000 files per minute, while intermittent-encryption variants like BlackBasta and Rorschach can reach 50,000 files per minute by targeting only partial file segments \cite{Townsend2024,RaaS,Lyons2022,putrevu2024comprehensive}.
    Underscoring the severity of these attacks, a Splunk study found that Babuk, LockBit, and REvil were able to rapidly encrypt a 98{,}000-file corpus (totaling 53\,GB), demonstrating their speed and destructive potential~\cite{Splunk2022}. 
    Evidently, there is an urgent need for solutions to detect and mitigate ransomware~\cite{moody_ransomware_2025}.

    \begin{figure}[ht]
        \centering
        \includegraphics[width=0.65\linewidth]{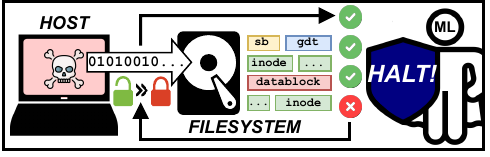}
        \caption{\textbf{High-level view of our approach}---as ransomware encrypts host files, SHIELD detects it in real time from filesystem metrics via closed-loop ML and promptly halts the disk}
        \label{fig:teaser}
    \end{figure}

    Ransomware's effectiveness, low barrier to entry, and covert payment channels make it a preferred tool for inflicting financial, reputational, and data-loss damage. 
    Despite defenses such as API-call monitoring, hardware performance counters, trap files, registry tracking, and network anomaly detection, organizations remain exposed since attacks can span network, OS, and hardware layers~\cite{Chen2017, Anand2023, Ganfure2023, Gomez-Hernandez2022,kotov2023,beaman2021ransomware,anand2023rtr}. 
    On-host tools rely on metrics that become untrustworthy once the OS is compromised and often lack filesystem-level granularity. 
    Moreover, none of these defenses can execute inside or adjacent to a storage controller since they depend on host-resident instrumentation that is unavailable in an embedded controller environment. 
    Cloud solutions, while isolated from host tampering, move data and security operations offsite, conflicting with organizational policies~\cite{sohail2023data}. 
    As ransomware variants grow increasingly sophisticated and embedded storage devices gain computational capabilities, there is both an urgent need and a practical opportunity for a low-level, tamper-proof, and host-independent detection solution that operates within the storage controller datapath.
     
   \begin{figure}[h]
        \centering
        \includegraphics[width=0.9\linewidth]{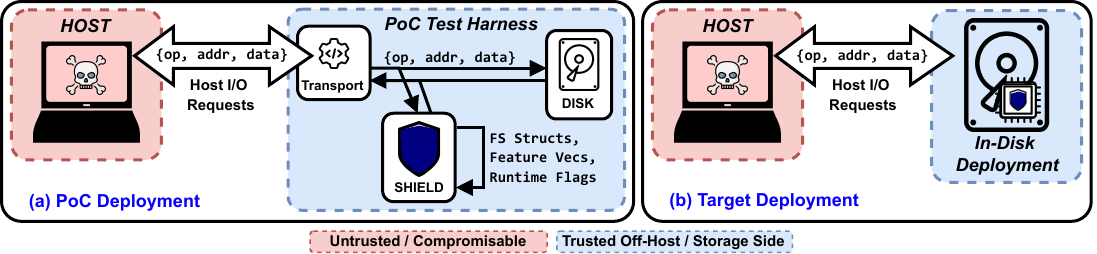}
        \caption{\textbf{Prototype vs Target Deployment.} Left: PoC evaluation harness for trace collection and validation of SHIELD (this work). Right: controller-integrated SHIELD deployment with in-line acquisition and enforcement, removing OS dependence in the monitoring path (future work).}
        \label{fig:trust_poc}
    \end{figure}
    Motivated by these challenges, we propose SHIELD, a secure, host-independent metric-acquisition framework that exposes fine-grained filesystem activity from below the host OS.
    SHIELD parses on-disk filesystem structures to produce tamper-resistant telemetry that cannot be suppressed or forged by a compromised kernel, enabling robust detection even under partial or intermittent encryption.
    By training machine learning models on these off-host filesystem metrics, our system accurately distinguishes benign from malicious disk behavior, differentiates between ransomware strains, and retains detection efficacy on previously unseen variants. Figure~\ref{fig:teaser} provides a high-level overview of our approach.
    The framework's streaming, bounded-compute design targets FPGA/ASIC controller integration, aligning with the growing trend toward in-controller intelligence in computational storage and smart SSD architectures.
    To evaluate feasibility and collect repeatable traces, we instantiate this framework in a prototype test harness that exposes standard storage disk read and write commands via a network block device interface. %
	Figure~\ref{fig:trust_poc} depicts the prototype's trust boundaries and contrasts the evaluation harness with the intended deployment, where the same metric-acquisition logic provides an off-host trust anchor at the storage level. We make the following contributions:

    \begin{itemize}[noitemsep,nolistsep,leftmargin=*]
        \item \textbf{Novel Off‐Host Metric Acquisition Framework:} We introduce a low‐level data acquisition pipeline that logs filesystem‐specific features (e.g., inodes, data blocks) independent of the OS.
        \item \textbf{Validation of Metric Utility for Detection:} We validate these metrics through comprehensive ML experiments, training both binary (benign vs.\ malicious) and multiclass (ransomware family) classifiers, and show that they enable high-accuracy discrimination of ransomware behavior ($97.29\%$ binary accuracy).
        \item \textbf{Zero-shot Detection \& Mitigation:} We show that models trained on these metrics can effectively identify and stop new, previously unseen ransomware samples and families in real time with minimum file loss (often only a few hundred bytes) while maintaining high accuracy.
        \item \textbf{Validation of Hardware Feasibility:} SHIELD is designed for controller-level deployment (FPGA or disk-controller ASIC), enabling fully off-host detection with inherent isolation. Hardware-only ablations show models trained without transport layer indicators retain high accuracy ($95.97\%$), confirming viability for designs that retire kernel and network-based metrics.
    \end{itemize}

    This paper proceeds as follows: 
    Section~\ref{sec:background_related} reviews background and situates our work; 
    Section~\ref{sec:threat_model} states the threat model and security goals; Section~\ref{sec:architecture} describes SHIELD's architecture and design; Section~\ref{sec:implementation} covers implementation and experimental setup; Section~\ref{sec:results} evaluates results and SHIELD's performance; 
    Section~\ref{sec:discussion} considers implications and practicalities; 
    Section~\ref{sec:conclusion} concludes.

\section{Background \& Related Works} \label{sec:background_related}

    \subsection{Ransomware Primer}
    From early threats (e.g., PC Cyborg Trojan) to modern campaigns, ransomware has matured to adopt strong cryptography, data exfiltration, and multi-stage extortion; outbreaks like WannaCry and NotPetya proved global reach, while Ransomware-as-a-Service (RaaS) commoditized attacks for low-skill actors~\cite{CrowdStrike2023, Microsoft2021}. Attack vectors include phishing, exploit kits, and remote desktop protocol flaws, with encryption of critical files in minutes. Consequences of crypto-ransomware range from data loss and downtime to life-critical disruptions; in embedded systems, dedicated hardware watchdogs that monitor untrusted components from a separate trusted core have proven effective for malware defense, motivating analogous approaches at the storage layer~\cite{Kadiyala2020TECSLAMBDA}. 

    Despite using varied tactics, all ransomware drive systematic disk activity, which includes intensive reads/writes to data blocks, file indexes, and file metadata, often alongside CPU/GPU spikes, abnormal power draw, and rapid file access behaviors~\cite{shieldfs,Dimov2020}. We focus our study on leveraging observable activity within deep filesystem features for driving detection.

    \begin{table}[h]
        \centering
        \caption{Conceptual mapping of major filesystem abstractions with core concepts across \texttt{NTFS}, \texttt{ext4}, and \texttt{APFS}, used to motivate SHIELD's modular backend parsing across filesystems. Here, mappings are representative examples and each filesystem includes additional structures not shown (journals, metadata copies, etc.).}
        \label{tab:fs_feats}
        \resizebox{0.9\linewidth}{!}{
        \begin{tabular}{rccc}
            \toprule
            \textbf{Filesystem Concept} & \textbf{\texttt{NTFS}} & \textbf{\texttt{ext4}} & \textbf{\texttt{APFS}}  \\  
            \cmidrule(lr){1-1} \cmidrule(lr){2-4} 
            \textbf{Directory Indexing}  & B-Tree indexed Directories  & H-Tree indexed Directories   &   B-Tree indexed Directories  \\
            \textbf{Global Metadata}     & Master File Table (MFT)   &  Superblock, Group Descriptors  &  Metadata B-Trees  \\
            \textbf{Per-File Metadata}   & MFT Records &  Inodes           &  Object Records            \\
            \textbf{Data Placement}      & Extents  &  Extents            &  Extents      \\  
            \textbf{Allocation Unit}     & Clusters           &  Blocks              &  Blocks      \\
            \textbf{Space Tracking}     & Bitmaps           &  Bitmaps              & Bitmaps    \\
            \bottomrule
        \end{tabular}
        }
    \end{table}
    
    \subsection{Filesystems in Context}
        During a ransomware attack, encryption or exfiltration of valuable data depends on the underlying filesystem. Unlike OS-level telemetry, since all changes and requests pass through the disk hardware itself, this layer cannot be obfuscated or spoofed. Regardless of attack strategy, timing, or hardware use, the ransomware must access, rename, or overwrite files to execute a successful attack, making the filesystem a critical layer for identifying anomalous behavior.
        Modern filesystems such as \texttt{ext4} (Linux), \texttt{NTFS} (Windows), and \texttt{APFS} (macOS) serve as the organizing layer between raw block storage and structured file lists. They manage naming, placement, metadata, and allocation through a range of robust, metadata-rich structures. These structures are not only vital for correctness and recovery but also provide a premise for forensic and behavioral modeling~\cite{fs_forensics}. Table~\ref{tab:fs_feats} outlines structural parallels across major filesystems, illustrating how semantically similar operations manifest differently depending on platform. Each major component within the filesystem plays a role in characterizing system behavior:

        \begin{itemize}[noitemsep,nolistsep,leftmargin=*]
            \item \textbf{Directory Structures \& Global Metadata} maintain hierarchical features and are frequently updated during file creation, movement, or renaming.
            \item \textbf{Per-File Metadata} tracks individual file properties such as timestamps, ownership, and data block locations, and are often mutated during ransomware activity.
            \item \textbf{Data Placement} is governed by extent-based (i.e., offset and size) allocation in all three filesystems, enabling efficient storage of large contiguous files.
            \item \textbf{Allocation Units} are the fundamental units of data organization made up of clusters or blocks. 
            \item \textbf{Space Tracking mechanisms} are generally implemented using bitmaps and reflect overall disk usage and allocations or deallocations during bulk file operations.
        \end{itemize}
        
        Because these exist on the disk beneath the OS kernel, they provide a tamper-resistant and generalizable source of telemetry. Even in the case of host compromise, meaningful file operations are impossible without inducing observable filesystem-level effects. Our framework leverages this invariance to extract semantically aligned metrics across platforms. For instance, although \texttt{NTFS} represents per-file metadata using MFT records and \texttt{APFS} uses object records, these structures serve analogous roles and exhibit comparable mutation patterns during file creation, traversal, renaming, and overwrite, enabling cross-OS behavioral characterization from a common event schema~\cite{fs_forensics}.

\begin{figure}[h]
            \centering
            \includegraphics[width=0.55\linewidth]{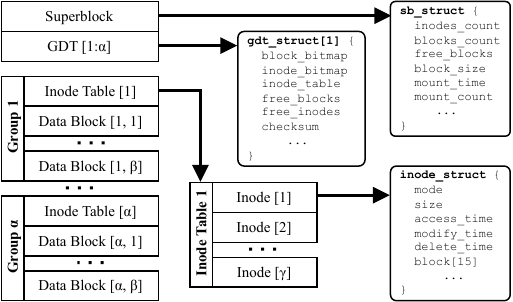}
            \caption{\textbf{A simplified \texttt{ext4} filesystem} with \(\alpha\) block groups, each containing \(\beta\) data blocks and \(\gamma\) inodes. Each feature indicates metrics which SHIELD can parse, monitor, and log.}
            \label{fig:ext4}
        \end{figure}

        Although SHIELD's backend is modular and new filesystems can be supported by implementing a parser that maps on-disk structures to the same feature categories, we focus on \texttt{ext4} due to its wide deployment and well-documented, open implementation, which make it a practical baseline for validating metric utility. Figure~\ref{fig:ext4} summarizes a simplified \texttt{ext4} hierarchy and highlights the core on-disk structures we instrument, which mediate metadata updates, allocation, and data placement. This storage-side, filesystem-semantic vantage point is distinct from, but complementary to, ransomware defenses at the application, kernel, or device layers; we next survey those approaches and discuss their limitations under host compromise.

    \subsection{Existing Defenses Against Ransomware}
        The evolving malware landscape now demands cross-layer detection and mitigation, shifting from signature-based antivirus to machine learning, behavioral analytics, and hardware-assisted methods to enhance defensive capabilities~\cite{razaulla2023age,CrowdStrike2023,9072101, 9474050}. 
        Common ransomware defenses span the application, kernel, and hardware levels, with some integrated into co-compute hardware or attempting to leverage surface-level filesystem features.

        \subsubsection{Orthogonal Defenses}
        Application-layer methods detect tampering via hashes/metadata but lack real-time response and are easily evaded; file decoys and signature AV likewise offer limited, update-heavy, tamper-prone protection~\cite{Murphy2024, signature}. Machine learning methods reduce manual rules yet still depend on host-resident data from the application or kernel layer~\cite{CybotsAI2024, Microsoft2021}. Kernel monitors expose file and network activity but add software overhead and fail once the host is compromised; even combined application--kernel--ML approaches remain on-host making them untrustworthy post-compromise~\cite{ranker}. Hardware monitors (processor traces, performance counters, trusted platform modules) improve runtime integrity but miss filesystem-specific attacks~\cite{trace}. Interposer-based approaches passively observe coarse logical block address (LBA) level or time-windowed I/O statistics but miss touches and mutations to critical filesystem structures in real-time~\cite{lophi}.
        Off-host options provide isolation but fall short: sandboxing is post-mortem, not real-time; cloud offloading often violates data policies and lacks filesystem insight~\cite{malone2011hardware, Anand2023, CrowdStrike2023}.
    
        \subsubsection{Filesystem Aligned Defenses}
        Few prior methods use filesystem-level metrics for ML-assisted ransomware defense in/near storage, in alignment with SHIELD. IBM FlashSystem runs hardware-assisted ML on a device-local Linux SoC for real-time protection~\cite{pletka_building_2024}. Reategui \emph{et al} detect attacks from kernel-generated storage traces within computational storage (i.e., near-data processing)~\cite{reategui_generalizability_2024}. Wang \emph{et al} propose a cloud detector using LBA mappings and I/O context (e.g., read-after-write), plus entropy features~\cite{wang_ransom_2024}. Table~\ref{tab:methodology_comparison} contrasts post-payload protections across in-storage and other approaches with SHIELD; key differences between these methods follow.

\newcommand*\rot[1]{\hbox to1em{\hss\rotatebox[origin=br]{-30}{#1}}}
\newcommand{\tmidrule}{\cmidrule(lr){1-2} \cmidrule(lr){3-9}}

\begin{table}[!t]
    \centering
    \caption{Comparison of Ransomware Detection and Mitigation Architectures Post Payload Execution.}
    \label{tab:methodology_comparison}
    \resizebox{0.55\linewidth}{!}{%
    \setlength\extrarowheight{-3pt}
    \begin{tabular}{lcccccccc}
    \toprule
    \textbf{\hspace{0.9cm} Architecture}   &
    {\textbf{ \thead{Metric \\ Acquisition Level}}} & 

\multicolumn{7}{c}{\textbf{\thead{Architecture \\ Security Parameters}}} \\
\tmidrule
& & 
    
    \rot{\textbf{ \thead{FS-Specific Features}}  }&
    \rot{\textbf{ \thead{Tamper Resistant}}} & 
    \rot{\textbf{ \thead{Host Independent}}} & 
    \rot{\textbf{ \thead{Hardware Assisted}}} & 
    \rot{\textbf{ \thead{Data Integrity Centric}}}&
    \rot{\textbf{ \thead{Real-time Mitigation}}}& 
    \rot{\textbf{ \thead{Evasion Proof}}} \\ 
    \tmidrule
    File Integrity Monitoring~\cite{Murphy2024}                      & Application    & \Circle      & \Circle      & \Circle      & \Circle       & \CIRCLE     & \Circle         & \Circle            \\ \tmidrule
    Decoy Files~\cite{putrevu2024comprehensive,Gomez-Hernandez2022}  & Application    & \Circle      & \Circle      & \Circle      & \Circle       & \CIRCLE     & \CIRCLE         & \Circle            \\ \tmidrule
    Signature-based Antivirus~\cite{signature}                       & Application    & \Circle      & \Circle      & \Circle      & \Circle       & \Circle     & \LEFTcircle     & \Circle            \\ \tmidrule
    AI/ML-based Detection~\cite{10521643}                            & App/Kernel     & \Circle      & \Circle      & \Circle      & \Circle       & \CIRCLE     & \CIRCLE         & \Circle            \\ \tmidrule
    Process Monitoring~\cite{ranker}                                 & Kernel         & \Circle      & \Circle      & \Circle      & \Circle       & \CIRCLE     & \CIRCLE         & \Circle            \\ \tmidrule
    Hardware Traces~\cite{trace}                                     & Hardware       & \Circle      & \CIRCLE      & \Circle      & \CIRCLE       & \Circle     & \CIRCLE         & \CIRCLE            \\ \tmidrule
    Performance Counters~\cite{malone2011hardware}                   & Hardware       & \Circle      & \CIRCLE      & \Circle      & \CIRCLE       & \Circle     & \CIRCLE         & \Circle            \\ \tmidrule
    Sandboxing~\cite{9965664}                                        & Off-host       & \Circle      & \CIRCLE      & \CIRCLE      & \Circle       & \Circle     & \Circle         & \Circle            \\ \tmidrule
    Cloud-based Security~\cite{Microsoft2021}                        & Off-host       & \Circle      & \CIRCLE      & \CIRCLE      & \Circle       & \Circle     & \Circle         & \Circle            \\ \tmidrule
    NVMe-oE SSD Logging~\cite{rssd}                                  & Hardware       & \Circle      & \CIRCLE      & \CIRCLE      & \CIRCLE       & \Circle     & \Circle         & \CIRCLE            \\ \tmidrule
    Filesystem-aware~\cite{shieldfs,Dimov2020}                       & Kernel         & \Circle      & \Circle      & \Circle      & \Circle       & \CIRCLE     & \CIRCLE         & \Circle            \\ \tmidrule
    Hardware Interposer~\cite{lophi}                                 & Hardware       & \Circle      & \CIRCLE      & \CIRCLE      & \CIRCLE       & \Circle     & \Circle         & \CIRCLE            \\ \tmidrule
    In-Storage Solutions~\cite{gagulic_ransomware_2023, pletka_building_2024, wang_ransom_2024}                           
             & Hardware       & \Circle  & \CIRCLE      & \CIRCLE      & \LEFTcircle   & \CIRCLE     & \CIRCLE         & \CIRCLE            \\ \tmidrule
    \textbf{\textsf{SHIELD}} (This Work)                             & Hardware       & \CIRCLE      & \CIRCLE      & \CIRCLE      & \CIRCLE       & \CIRCLE     & \CIRCLE         & \CIRCLE            \\ \bottomrule \\
    
    \multicolumn{9}{c}{\Circle\hspace{0.5em}= No Support \qquad \LEFTcircle\hspace{0.5em} = Partial Support \qquad \CIRCLE\hspace{0.5em} = Full Support}
    \end{tabular}
    } 
    \end{table}

    \subsection{SHIELD in Comparison}
    Compared to the filesystem aligned defenses, SHIELD is distinct in that it's metric collection module can be driven  below the OS level, and does not require computational storage or co-processors/SoCs for software/firmware based calculation when implemented on disk hardware or controller ASICs. Furthermore, existing near-storage defenses rely on coarse LBA patterns and time-windowed I/O statistics, which blur metadata vs. data operations and delay detection (often seconds–minutes)~\cite{HIRANO2022301314,wang_ransom_2024}. LBA offers only location coarse-ness, further complicated by formatting regularities and adjacency of inode tables and data blocks in memory. SHIELD instead performs filesystem-aware parsing and action-based aggregation, exposing structure-specific activity (e.g., inode-table vs. data-block writes) and detecting intermittent encryption in as few as two disk actions; with millisecond-scale device latencies, this enables sub-second response while avoiding trust in host logs. Moreover, many in-storage require sequential co-processing, they favor low-complexity features~\cite{wang_ransom_2024}; in comparison, SHIELD targets FPGA/ASIC, allowing for parallelizing of parsing and entropy to cut overhead and reduce optimization effort.
                  
\vspace{-0.7em}
\section{Threat Model} \label{sec:threat_model}

    \subsection{System Model, Trust Boundaries}
        We consider a host system that issues standard block I/O requests to a protected storage volume. SHIELD exists within the storage access path and observes or initiates commands on the same block-level traffic and on-disk filesystem structures used to realize file operations. We assume the host and its operating system are untrusted. The off-host acquisition/decision component (target deployment: in-controller logic, FPGA, or storage-side appliance) is treated as trusted and physically protected. The host interacts only with a standard block-device interface (i.e., address, size, and read or write operation) and it has no channel to modify SHIELD's logging or decision logic. The resulting trust boundary, which includes the untrusted host OS versus trusted storage-side acquisition and enforcement is depicted in Figure~\ref{fig:trust_poc}.

    \subsection{Adversary Model}
        We assume an adversary with full compromise of the host OS, including the ability to terminate or tamper with on-host security software, inject kernel code, manipulate system calls, and modify host-resident logs. The adversary may attempt evasion via intermittent or partial encryption, multithreading, and I/O shaping.
        We assume the adversary does not have physical access to the off-host device and cannot reprogram its firmware/bitstream or perform supply-chain insertions. Side-channel attacks against the off-host device and physical bus probing are out of scope.

    \subsection{Security Goals \& Scope}
        SHIELD aims to (1) provide tamper-resistant telemetry by acquiring filesystem-aware metrics below the host OS, and (2) enable real-time detection and mitigation by halting further writes to the protected volume once ransomware-like behavior is detected, thereby bounding file loss. SHIELD does not guarantee availability under a powerful denial-of-service adversary; an attacker may disrupt access to the protected volume, but doing so prevents the attack's encryption progress rather than enabling stealthy corruption. SHIELD also does not attempt to prevent initial host compromise, and it protects only storage accesses that traverse the monitored path.

    \subsection{Prototype vs. Target Deployment} \label{sec:arch_poc_vs_target}
        Our evaluation uses a prototype test harness that routes I/O through a Linux network block device transport interface to (a) collect repeatable traces and (b) validate equivalence between software-visible I/O and off-host metrics. This does not define the trust boundary of the intended system. The target deployment places SHIELD's acquisition and decision logic within or adjacent to the storage controller (FPGA/ASIC/SoC), eliminating dependence on any general-purpose OS in the monitoring path. Importantly, even in the PoC, the compromised host cannot fabricate the underlying on-disk filesystem transitions that SHIELD derives at the storage interface; at most, the host can delay or deny its own access to the monitored volume.

    \subsection{Off-Host Compromise}
        Similar to existing storage-side defenses, our security claims assume the storage-side monitoring component is physically protected and not compromised. If an attacker compromises the off-host appliance in the PoC, they could suppress logging or mitigation; this is exactly why the target design integrates the acquisition and enforcement logic into the storage controller/FPGA boundary where host compromise cannot reach it. We treat hardening and attestation of an off-host appliance OS as orthogonal engineering, and we eliminate it in the intended in-controller deployment, which requires no OS or compute device.

\section{System Architecture} 
    \label{sec:architecture}

    \begin{figure*}[!tp]
        \centering
        \includegraphics[width=0.95\textwidth]{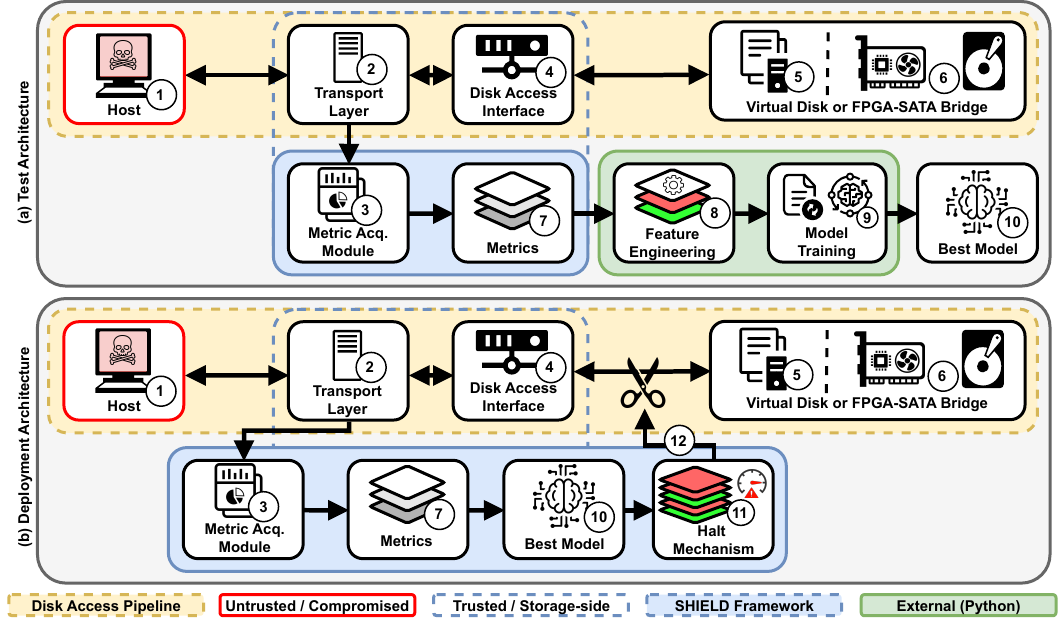}
        \caption{Architectural overview and flow of SHIELD components.}
        \label{fig:arch}
    \end{figure*} 

    \subsection{Overview \& Design Goals}
        \label{sec:arch_overview}
        
        Aligned with the threat model in Section~\ref{sec:threat_model}, we design SHIELD as a low-cost, modular off-host metric-acquisition framework that remains effective under full host compromise. The core objective is to obtain tamper-resistant, filesystem-aware telemetry from below the host OS, enabling reliable ransomware detection when on-host logs, sensors, and security agents can no longer be trusted. Using this telemetry, we demonstrate end-to-end feasibility by (i) collecting low-level disk and filesystem metrics, (ii) training supervised detectors on action-windowed features, and (iii) deploying the selected binary model in a closed-loop configuration that triggers write-blocking mechanisms to bound file loss when malicious behavior is detected.
        
        To evaluate the utility of the framework under controlled and repeatable conditions, we instantiate it in two complementary configurations. First, a \textbf{Test Architecture} generates a large labeled corpus by logging every observed disk action and aggregating actions into overlapping windows for training. Second, a \textbf{PoC Deployment} uses the best-performing binary model in an evaluation harness to perform real-time classification, mitigation, and data loss measurements. Throughout, our design choices favor hardware compatibility: the acquisition, feature generation, and inference components are structured so they can be migrated into a storage-side FPGA/ASIC pipeline for full isolation and inherent tamper-resistance (Section~\ref{sec:hw_cons}).
    
    \subsection{Overall System Design}
        \label{sec:arch_sysmodel}
        
        Figure~\ref{fig:arch} summarizes the system at a high level. Figure~\ref{fig:arch}.(a) shows the \textbf{Test Architecture}, which streams per-action telemetry and logs predefined features to build training vectors. Figure~\ref{fig:arch}.(b) shows the \textbf{Deployment Architecture}, which invokes a pre-trained binary classifier for online use with aggregated action-based vector inputs to enable real-time detection and enforcement.
        
        SHIELD is organized around a stable, storage-side acquisition pipeline that is shared across both configurations: disk I/O is observed at the disk-access boundary \circled{4}, interpreted by the metric acquisition and parsing logic \circled{3}, and emitted as per-action metric vectors \circled{7} that feed either offline feature engineering \circled{8} (test) or online inference \circled{10} (deployment). Consistent with our threat model, the host OS is treated as untrusted and interacts only through a standard block-device interface; all acquisition and enforcement logic resides within the trusted side of the boundary.
        
        We explicitly distinguish the framework from the prototype harness. The framework comprises filesystem parsing, event attribution, metric-vector generation, and action-window aggregation. The prototype test harness (i.e., network storage and block device layer) provides a convenient transport and orchestration mechanism for experiments. This transport layer is not required by the intended design and does not define the trust boundary: in the target deployment, the same acquisition and enforcement logic is integrated at the storage/controller boundary without reliance on a general-purpose OS or network block device interface (see Figure~\ref{fig:trust_poc} and Section~\ref{sec:arch_poc_vs_target}).

    \subsection{Metric Acquisition Framework} \label{sec:metric_framework}
        At the core of SHIELD is the metric acquisition module \circled{3}, a multilayer mechanism that observes storage activity below the host OS and derives filesystem-aware telemetry from on-disk state. Implementation level details for this module are given in Section~\ref{sec:acq_flow}. In our \texttt{ext4} instantiation, the framework parses filesystem structures \circled{5} \circled{6} directly from disk and optionally combines them with interface-level I/O statistics \circled{2} to produce a unified view of each access.
        The framework emits a multidimensional metric vector \circled{7} for every observed action, capturing the operation type (read/write), the affected filesystem structure,
        and associated per-file metadata and content-level signals
        ~\cite{ext4}. Collected metrics are summarized in Section~\ref{sec:implementation_Features}. As the filesystem changes, SHIELD continuously maintains and updates a live catalog of relevant structures through two parsing phases. The vectorized information is then aggregated for inference inputs.
    
        \subsubsection{Active vs. Passive Parsing}
            SHIELD derives host-independent on-disk events in two complementary phases. \textit{Active Parsing} runs at disk initialization prior to host access by querying global filesystem metadata to bootstrap a live catalog of inodes and their associated data block mappings. \textit{Passive Parsing} operates during normal disk use, monitoring subsequent reads and writes and attributing each action to the filesystem structure it touches. For each access, the framework attaches metadata describing the impacted structures and any observed mutations, enabling downstream feature vector aggregation and model training or real-time classification.
             
        \subsubsection{Action-based Windows}
            
            The metric acquisition framework aggregates telemetry at the granularity of a single action. We define an action as one storage I/O request observed at the disk interface. This equates to a single read or write operation characterized by an operation type (R/W), an LBA range, a request size, and the (optional) associated payload. 
            For training and online inference, SHIELD groups consecutive actions into action-based windows with configurable window sizes and overlaps (further detailed in Section~\ref{sec:windows}).

    \subsection{Test Architecture \& Prototype Harness}
        \label{sec:test_arch} 
        
        Within the \textbf{Test Architecture}, SHIELD records a stream of disk actions and emits a per-action metric vector \circled{7}. These logs are then externally labeled using the ground-truth of the workload being executed and aggregated into overlapping action-based windows during feature engineering \circled{8}. This produces a supervised training corpus that is robust to execution-time variability because each sample corresponds to a fixed number of disk operations.
        
        To collect repeatable traces and keep the monitored host logically isolated from the acquisition stack, our prototype employs an network block device based transport harness. This configuration provides a convenient mechanism to route standard block I/O to the off-host acquisition framework for logging and dataset generation. Moreover, the transport harness is not a required component of SHIELD's security boundary or long-term design; it is used here solely as an evaluation vehicle for controlled trace collection as outlined in Figure~\ref{fig:trust_poc} and discussed further in Section~\ref{sec:trust_discuss}.
        
        The resulting dataset is used for model development \circled{9}. We train and evaluate two classifier types to assess metric utility under different objectives: a \textbf{binary} classifier that distinguishes benign versus malicious behavior (used for real-time detection), and a \textbf{multiclass} classifier that predicts ransomware families to determine if metrics are capable of characterizing strain-specific access patterns. We perform standard hyper-parameter tuning across window configurations and select the best-performing binary model \circled{10} for online deployment.
    
    \subsection{Real-Time Detection \& Enforcement Interface}
        \label{sec:deploy_arch}
        
        After training, the selected binary classifier is integrated into a closed-loop detection path. SHIELD continuously aggregates actions into windows matching the training configuration and submits each window as a feature vector to the classifier. The classifier outputs a `maliciousness' score, which drives a halt mechanism \circled{11} and enforcement policy \circled{12} intended to bound damage by stopping further encryption as early as possible.
        
        \subsubsection{Action-triggered decision intervals}
            Decisions are made at fixed action thresholds (e.g., every $W$ actions for a window size $W$), rather than at fixed time intervals. This design makes detection latency comparable across heterogeneous systems and resists evasion strategies that slow down or sparsify encryption, since an attacker must still perform storage actions to encrypt files.

        \subsubsection{Decision Smoothing \& Enforcement Semantics}
            To suppress transient spikes, the PoC smooths predictions with a rolling window of the last five inferences. If malicious votes exceed a fixed threshold, SHIELD fails closed at the disk-access interface \circled{12} by blocking subsequent writes to the protected volume. The host sees normal I/O write failures, which prevents additional ciphertext from reaching disk. This lightweight policy is sufficient to demonstrate closed-loop enforcement and bound pre-halt corruption (Section~\ref{sec:results}).

    \subsection{Filesystem Backend Modularity}
        \label{sec:fs_modularity}
        
        We evaluate SHIELD using \texttt{ext4} because it is widely deployed, openly specified, and well documented, making it a practical foundation for validating correctness and end-to-end feasibility. Importantly, SHIELD is not tied to \texttt{ext4}. The metric acquisition framework is backend-agnostic: it interprets disk activity through a common, filesystem-aware event schema, allowing different filesystems to be supported by swapping the parsing backend while preserving the same downstream feature aggregation, learning, and mitigation pipeline.
        To motivate this portability, Table~\ref{tab:fs_feats} summarizes the approximate correspondence between major on-disk abstractions across \texttt{NTFS}, \texttt{ext4}, and \texttt{APFS}. Consistent with these correspondences, we have implemented an additional \texttt{NTFS} backend using \texttt{NTFS-3G} that maps \texttt{NTFS}-specific structures into the same schema used by our \texttt{ext4} backend~\cite{ntfs-3g}, demonstrating modularity and feasibility in supporting non-\texttt{ext4} filesystems.
        
        In a controller-integrated deployment, this modularity enables the filesystem parser to remain an interchangeable component behind a stable interface, supporting multi-filesystem environments without redesigning the learning or enforcement mechanisms. Implementation details of the \texttt{ext4} backend and the shared acquisition pipeline are provided in Section~\ref{sec:implementation}.

\subsection{Hardware Feasibility by Design}
\label{sec:hw_cons}

SHIELD performs storage-side ransomware detection so that telemetry and enforcement remain outside the host’s trust boundary. Although we evaluate it in a software harness for controlled experiments, the pipeline is designed to map directly to an FPGA/ASIC storage-controller implementation. In deployment, acquisition and mitigation run in or alongside the controller, removing dependence on host software and remaining tamper-resistant under full OS compromise. We emphasize feasibility through the following design constraints:

\begin{itemize}[noitemsep,nolistsep,leftmargin=*]
    \item \textbf{Streaming, per-action processing:} Normal operation processes each block I/O request as it arrives, avoiding unbounded scans or expensive global recomputation. State is maintained as incremental, bounded updates to cached filesystem metadata and rolling aggregates.
    \item \textbf{Bounded per-action compute:} Telemetry extraction uses simple indexed lookups and table-driven parsing, making the acquisition path compatible with controller-grade memory budgets.
    \item \textbf{Synthesizable inference:} Models are trained off-device (Figure~\ref{fig:arch}) and exported as compact arithmetic/branching routines suitable for hardware synthesis without external dependencies.
    \item \textbf{Transport Harness in PoC:} The software network block device based transport layer within the evaluation harness is used for repeatable trace collection and closed-loop testing. A controller-integrated design attaches directly to the storage datapath (e.g., SATA/NVMe) and derives telemetry from observed commands and on-disk parsing. Consistent with this goal, we evaluate a hardware-only feature set (disabling transport-derived features) and retain high accuracy.
    \item \textbf{Modular hardware mapping:} SHIELD decomposes cleanly into (i) I/O access, (ii) filesystem attribution, (iii) rolling action-window aggregation, (iv) inference, and (v) logging/enforcement. This supports hierarchical mapping to FPGA fabric and migration into an ASIC. Filesystem support is abstracted behind the parser interface, enabling filesystem modularity.
    \item \textbf{Analytic feasibility and speed-independence:} Core operations such as metadata lookups, counter/ratio updates, feature generation, and classification are all streaming and parallelizable, allowing overlap with host I/O and metadata DMA. Because decisions are action-based rather than time-based, detection latency remains stable across device speeds, and controller-side integration removes the VM/transport overhead of the PoC (Section~\ref{sec:overhead}).
\end{itemize}

\section{Implementation \& Experimental Setup}\label{sec:implementation}
    Guided by the threat model, trust boundary,  and the proposed architecture, we implement the SHIELD PoC as a C-based, storage-side metric acquisition framework that derives filesystem-aware telemetry from the disk I/O stream and maintains a live on-disk filesystem catalog. This implementation serves two purposes: (i) it supports the test architecture by producing high-resolution, per-action logs for dataset generation and model training, and (ii) it supports the deployment architecture by aggregating actions into windows and invoking a pre-trained classifier in a closed loop for zero-shot detection and mitigation experiments. 
    For testing, we embed both architectures within a test harness using a network block device-based transport interface for controlled trace collection; however, the core logic is independent of the transport interface and designed to map to storage/controller deployments (Section~\ref{sec:hw_cons}).

    \begin{figure}[!t]
        \centering
        \includegraphics[width=0.7\linewidth]{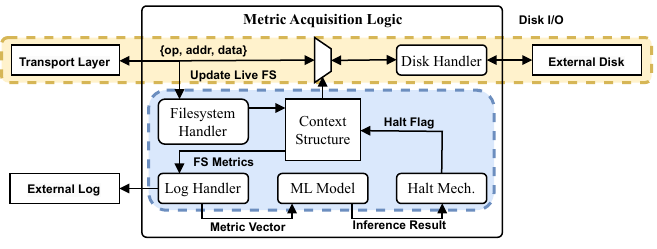}
        \caption{Internal modules and data flow in the SHIELD prototype implementation.}
        \label{fig:exec_flow}
    \end{figure}
    
    \begin{figure}
        \centering
        \includegraphics[width=0.9\linewidth]{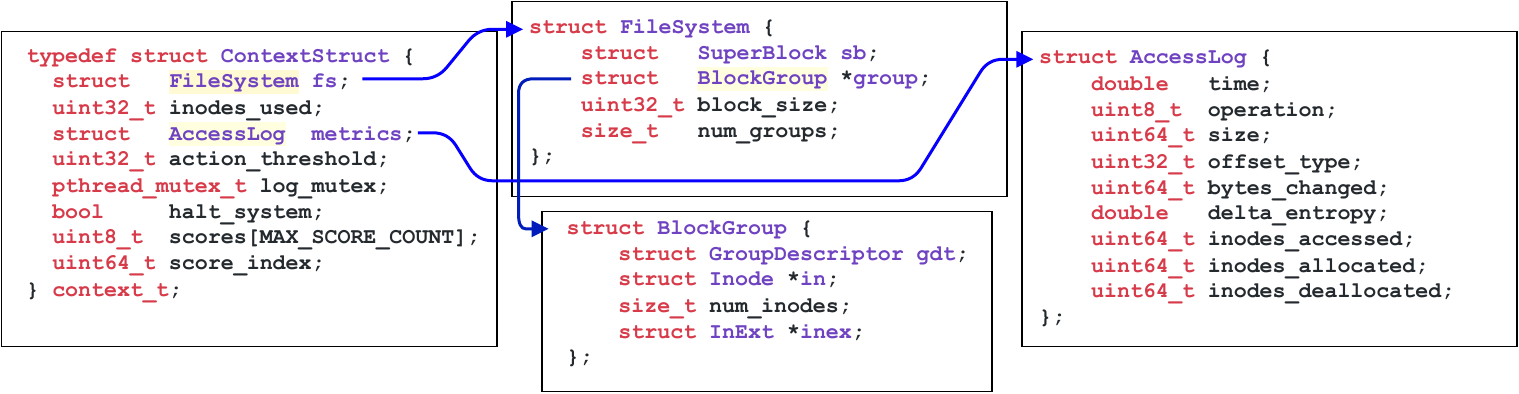}
        \caption{Shared hierarchical context structure used by all modules, including the live filesystem catalog, per-action metrics, and window-level buffers for classification.}
        \label{fig:context_struct}
    \end{figure}

    \subsection{Modules, Control Flow, \& Event Schema}\label{sec:acq_flow}

        Figure~\ref{fig:exec_flow} illustrates the control flow of our implementation. SHIELD is composed of a small set of modules that share a central context structure (Figure~\ref{fig:context_struct}) to maintain the live filesystem catalog, per-action state, and logging/aggregation buffers:
    
        \begin{itemize}[noitemsep,nolistsep,leftmargin=*]
            \item \textbf{Main Module} orchestrates initialization, configures mode of operation, interfaces with transport mechanisms, and manages high-level control (including enabling enforcement).
            \item \textbf{Context Structure} (Figure~\ref{fig:context_struct}) stores shared state, including the filesystem catalog, runtime flags, per-action metric fields, and window-level buffers used for aggregation and inference.
            \item \textbf{Filesystem Handler} parses on-disk \texttt{ext4} structures. At initialization it scans global metadata and inode tables to bootstrap mappings from file metadata to data-block locations; during runtime it incrementally updates this catalog as allocations and metadata mutations occur.
            \item \textbf{Disk Handler} abstracts the underlying storage backend, exposing a uniform read/write interface for both virtual disk images and physical disks.
            \item \textbf{Log Handler} constructs the per-action metric vector and either (i) emits one record per action for offline dataset generation (test mode), or (ii) aggregates actions into a window until the configured threshold is reached for online classification (deployment mode).
            \item \textbf{ML Handler} applies the selected binary classifier to each window-level feature vector and outputs a maliciousness score in $[0,1]$.
            \item \textbf{Halt/Enforcement Handler} applies a simple smoothing policy over recent classifier outputs and, when triggered, sets an enforcement flag that causes subsequent writes to be blocked.
        \end{itemize}

        In practice, each incoming disk I/O request from the (potentially compromised) host is treated as a single action (a single read or write command to the disk characterized by an operation type, an LBA range, a request size, and a payload for writes). For each action, the Filesystem Handler attributes the request to the relevant \texttt{ext4} structure(s) using disk offsets and the live catalog, updates catalog state when metadata changes are observed, and passes the resulting attribution to the Log Handler. In \textbf{test mode}, SHIELD logs a complete per-action metric vector (action threshold $=1$) to generate high-resolution traces for offline feature engineering and model training. In \textbf{deployment mode}, SHIELD aggregates consecutive actions into a window (e.g., threshold $=20$) and forwards the aggregated feature vector to the ML Handler for inference. The Enforcement Handler maintains a rolling history of recent window-level decisions; if malicious evidence exceeds a configured threshold, an enforcement flag is asserted and the Disk Handler blocks subsequent write requests, preventing further encryption progress.

    \subsection{Storage Backend \& Disk Setup}      
        We implement the SHIELD prototype to support both virtual and physical storage backends. Virtual disks provide near-native I/O performance and enable fast, repeatable experiments via cloning and snapshots for resets. Physical disk access is supported through an open-source FPGA-based SATA host bus adapter (HBA) on a Digilent XUPV5 board~\cite{AMD_UG347,Digilent_Virtex5}, which provides a hardware-realistic datapath and a convenient platform for validating storage-side observability. We extend this HBA (originally associated with Microsoft SIRC) to be OS-agnostic and accessible over Ethernet, allowing any host to interact with the disk through a standard block-device interface~\cite{MLE_ZynqSSE,sirc,groundhog}. Unless otherwise noted, experiments are run on virtual disks for scalability, and key behaviors are confirmed on the physical backend for consistency.
        
        For the virtual setup, we create a 15\,GB \texttt{ext4}-formatted baseline disk image populated with approximately 10\,GB of randomly generated files. The baseline image is generated with the \verb|mkfs.ext4| command.
        For each run, the baseline is cloned, attached to the test harness, and discarded afterward, ensuring that each execution exhibits comparable filesystem state while preserving identical metric-collection semantics. For physical testing, we replicate the same \texttt{ext4} layout onto a SATA HDD connected via the FPGA-SATA HBA and reformat/restore the drive between experiments.

        \begin{table}[h]
            \centering
            \caption{Disk Information for Ransomware Testing}
            \label{tab:disks}
            \resizebox{0.6\linewidth}{!}{
                \begin{tabular}{rccccc}
                \toprule
                \textbf{Disk} & \textbf{Util. / Capacity (MB)}   & \textbf{No. Files} & \textbf{Inodes Used} & \textbf{Data Blocks Used} \\
                \midrule
                1 & 9776.81 / 15360 & 22{,}486 & 22{,}510 & 2{,}502{,}864 \\
                2 & 9757.42 / 15360 & 20{,}833 & 20{,}845 & 2{,}497{,}901 \\
                3 & 9760.89 / 15360 & 21{,}782 & 21{,}794 & 2{,}498{,}790 \\
                4 & 9823.35 / 15360 & 21{,}960 & 21{,}972 & 2{,}514{,}778 \\ \hline
                \end{tabular} 
                }
        \end{table}
        
        To capture realistic variation in filesystem layout and utilization, we generate four distinct \texttt{ext4} disk images whose structural statistics are summarized in Table~\ref{tab:disks}. Here, Disks 1--3 are used to generate training data, and we hold out Disk 4 for zero-shot tests on ransomware samples and families not seen during training and test.

        \begin{table}[]
            \centering
            \caption{Collected features per disk action: \textbf{(a)} transport/interface counters observed on the trusted acquisition side and \textbf{(b)} filesystem-aware features derived by parsing on-disk \texttt{ext4} structures.}
            \label{tab:collected_features}
            \resizebox{0.6\linewidth}{!}{
            \setlength\extrarowheight{-3pt}
            \begin{tabular}{lll}
            \toprule
            \textbf{(a) Transport-Level Metrics} & \multicolumn{2}{c}{\textbf{(b) Filesystem-Level Metrics}}  \\  
            \cmidrule(lr){1-1} \cmidrule(lr){2-3}
            Reads Completed          &  Total Reads          & GDT Writes\\
            Reads Merged             &  Total Writes         & Inode Table Reads\\
            Sectors Read             &  Size                 & Inode Table Writes\\
            Time Spent Reading       &  Bytes Changed        & Data Block Reads\\
            Writes Completed         &  Sum Delta Entropy    & Data Block Writes\\  
            Writes Merged            &  Avg Delta Entropy    & Inodes Accessed \\  
            Sectors Written          &  Superblock Reads     & Inodes Allocated \\
            Time Spent Writing       &  Superblock Writes    & Inodes Deallocated \\  
                                     &  GDT Reads     & \\ \bottomrule
            \end{tabular}}
        \end{table}

    \subsection{Metrics Acquired}
    \label{sec:implementation_Features}
        
        At a high level, SHIELD bootstraps a consistent view of \texttt{ext4} metadata and initializes the logging pipeline. During runtime, for each observed disk action, the framework (i) attributes the request to the relevant filesystem structure(s) based on disk offsets and the live catalog, (ii) records the access and any detected mutations, and (iii) updates its local mirror of filesystem state. Pseudo-code for initialization and per-action logging is provided in algorithm~\ref{alg:acquisition} in Appendix~\ref{sec:algo}.
        
        For each disk action (read/write), SHIELD emits a metric vector consisting of 25 features consisting of 8 transport/interface counters and 17 filesystem-derived features (Table~\ref{tab:collected_features}). We additionally compute entropy-derived signals (Section~\ref{sec:entropy}), which contribute two summary statistics per action window (total and average change in entropy). All features are marshaled through the shared context structure (Figure~\ref{fig:context_struct}) and vectorized/recorded by the Log Handler.

        \subsubsection{Transport/Interface Metrics}
            Column (a) of Table~\ref{tab:collected_features} reports metrics available at the trusted acquisition boundary, such as read/write completion counts, merged requests, sectors transferred, and time spent in reads/writes. In our prototype harness these values are obtained from the trusted disk-access transport endpoint (rather than from the compromised host OS). We use these counters for two purposes: (i) as a sanity check that interface-observable activity is consistent with filesystem-attributed activity during trace collection, and (ii) as optional features in the PoC analogous to when SHIELD is used alongside other defenses. To align with the target controller-integrated deployment, we also evaluate hardware-only configurations that drop these transport-level features entirely and rely solely on filesystem-derived metrics (Section~\ref{sec:hardware_only}).
    
        \subsubsection{Core Filesystem-Level Metrics}
            Column (b) of Table~\ref{tab:collected_features} outlines the filesystem-level features derived by parsing on-disk \texttt{ext4} structures,  attributing each action to the structure it touches. These include aggregate read/write activity, request sizes, bytes changed, and counts of accesses to specific metadata regions as well as data blocks. Inode metadata is further analyzed to detect allocations and deallocations, which are recorded in the live catalog and surfaced as explicit features in the metric vector. Together, these features capture the structural footprint of file operations at the disk level, independent of host-resident instrumentation.

    \subsubsection{Entropy Feature}
        \label{sec:entropy}
        
        To complement the filesystem-derived metrics with a content-sensitive signal, we compute Shannon entropy over the byte distribution of written data blocks. Let $B_i$ denote the post-write byte sequence of a block after update $i$, and let $p_{i,b}$ be the empirical probability of byte value $b \in \{0,\dots,255\}$ in $B_i$. The Shannon entropy (in bits) is
        \begin{equation}
        \footnotesize
        \label{eq:shannon_entropy}
        H(B_i)
        = -\sum_{b=0}^{255} p_{i,b}\log_2 p_{i,b},
        \qquad
        p_{i,b}=\frac{n_{i,b}}{|B_i|},
        \end{equation}
        where $n_{i,b}$ is the number of occurrences of byte value $b$ in $B_i$, $|B_i|$ is the block size in bytes, and we adopt the standard convention $0\log_2 0 \triangleq 0$. For 8-bit data, $H(B_i)\in[0,8]$.
        
        We define the per-update entropy change (\emph{delta entropy}) as $\Delta H_i = H(B_i) - H(B_{i-1})$.
        Sustained positive shifts in entropy are a common signature of encryption-like transformations and provide an additional discriminator for ransomware behavior. To capture these trends, we aggregate $\Delta H_i$ over a sliding window of recent updates, where for a window starting at index $t$, $W_t \triangleq \{t,\, t+1,\, \dots,\, t+|W|-1\}$.
        This results in two summary statistics:
        (i) the \emph{summed delta entropy}, which captures the cumulative entropy change over the window,
        \begin{equation}
        \footnotesize
        \label{eq:sum_delta_entropy}
        S_{\Delta H}(t)
        = \sum_{i \in W_t} \Delta H_i,
        \end{equation}
        and (ii) the \emph{average delta entropy}, which measures the mean entropy change per update,
        \begin{equation}
        \footnotesize
        \label{eq:avg_delta_entropy}
        \overline{\Delta H}(t)
        = \frac{1}{|W|}\sum_{i \in W_t} \Delta H_i.
        \end{equation}

    \subsection{Window Aggregation \& Sample Labeling} \label{sec:window_labeling} \label{sec:windows}
        To convert the per-action metric stream into supervised learning samples, SHIELD groups consecutive disk actions into action-based windows of size $\mathit{W}$ with overlap $\mathit{O}$, resulting in window $(W,O)$. Each new window begins $\mathit{W}-\mathit{O}$ actions after the previous one, which increases temporal context while preserving responsiveness. We evaluate window sizes $(W)$ from 2 to 100 actions and overlaps $(O)$ from 0 to 50 actions. For each window, per-action features are aggregated (e.g., summed or averaged depending on the feature) to produce a single fixed-length feature vector.
        Each windowed sample is labeled as benign (0) or malicious (1) based on the executed workload, and windows derived from ransomware runs additionally carry a family identifier for multiclass classification. Table~\ref{tab:window_summary} (Section~\ref{sec:results}) summarizes the window configurations and the resulting sample counts.
        
        During online operation using the pretrained model, the same windowing mechanism is implemented via an action threshold parameter in the shared context structure, which specifies how many actions are accumulated before producing one inference sample (e.g., $W=$ 2, 20, or 100 actions). Smaller thresholds reduce detection latency but may reduce accuracy due to limited context; we quantify this tradeoff in our mitigation experiments by measuring both actions-to-detection and pre-halt file corruption on unseen ransomware strains.

    \subsection{Model Training, Export, \& Online Inference}
        \label{sec:model_training}
        
        We train several classifiers using an 80/20 train-test split and evaluate each model across a hyperparameter grid covering typical ranges for each algorithm. The full hyperparameter search grid is given in Table~\ref{tab:hyperparam_grids} within Appendix~\ref{sec:hyperparams}). Using these parameters, we reporting accuracy, precision, recall, and F1. The same pipeline supports both binary (benign vs.\ malicious) and multiclass (family identification) tasks, with the target label being the primary difference. We evaluate all classifier--hyperparameter--window--overlap combinations and produce a performance matrix from which the best configurations are selected. For deployment and real-time inference, we export the best-performing binary classifier to C using \texttt{m2cgen} and compile the generated inference code directly into the SHIELD runtime~\cite{m2cgen}. This enables real-time classification on windowed feature vectors without external ML dependencies, and preserves portability to future hardware implementations.
    
    \subsection{Chosen Software}
        \label{sec:chosen_software}
        
        Table~\ref{tab:software_lists} summarizes the workloads used to evaluate SHIELD. Our goal is to stress the metric-acquisition pipeline with (i) ransomware which exhibits diverse, real-world encryption behaviors and (ii) benign applications that generate heavy and heterogeneous disk activity.

            \begin{table}[h]
            \centering
            \caption{Software: (a) Ransomware,  (b) Benignware and (c) Unseen Strains/Families not used in training}
            \label{tab:software_lists}
                \resizebox{0.6\linewidth}{!}{
                \setlength\extrarowheight{-3pt}
                \begin{tabular}{rccc}
                \toprule
                 & \textbf{(a) Test Ransomware} & \textbf{(b) Test Benignware} & \textbf{(c) Unseen Ransomware}  \\  
                \cmidrule(lr){2-2} \cmidrule(lr){3-3} \cmidrule(lr){4-4}
                1. & Atomsilo      &  OBS          &  Akira  \\     
                2. & AvosLocker    &  VLC          &  Phobos  \\     
                3. & Babuk         &  7Zip         &  Inc  \\ 
                4. & Conti         &  Gimp         &  DragonForce  \\ 
                5. & GlobeImposter &  Eraser       &  Trigona  \\         
                6. & Intercobros   &  Kdenlive     &  HelloDown  \\  
                7. & Lockbit       &  Handbrake    &  Expiro  \\     
                8. & Makop         &  Veracrypt    &  LokiLocker  \\ 
                9. & MountLocker   &  Ultrasearch  &  Lynx  \\         
                10.& Fog           &  qBittorrent  &  CryLock  \\ 
                 \bottomrule
            \end{tabular}}
            \end{table}

        \subsubsection{Ransomware for Training and Test}
            We select ten ransomware families (Table~\ref{tab:software_lists}.a) that cover a range of behaviors relevant to disk-centric detection, including differences in file discovery and enumeration, file-type filtering, concurrency, and I/O intensity. Many families encrypt large sets of files by repeatedly reading plaintext blocks and overwriting them with high-entropy ciphertext, which produces characteristic patterns in metadata mutations (e.g., inode updates and allocation activity), data-block writes, and entropy shifts. We focus on families that are widely reported, operationally representative, and reliably executable in a controlled sandbox so that multiple repeated runs can be collected under consistent initial disk states.
            
        \subsubsection{Chosen Benign Applications}
            To reduce the risk of false positives, we include ten benign applications (Table~\ref{tab:software_lists}.b) with varied storage footprints and access patterns, spanning multimedia processing (OBS, VLC, Handbrake, Kdenlive), content creation (Gimp), search and indexing (Ultrasearch), and network-intensive I/O (qBittorrent). We also include high-write and disk-sanitization utilities (Eraser), archival/compression workloads (7Zip), and encryption software (Veracrypt). These programs are intentionally chosen because they can exhibit high write throughput and nontrivial metadata activity, providing realistic negative examples that are harder to distinguish from ransomware than lightweight office workloads.
        
        \subsubsection{Unseen ransomware for Zero-Shot Generalization}
            To evaluate generalization beyond the training distribution, we include ten additional ransomware strains (Table~\ref{tab:software_lists}.c) that are not used during training. These samples are zero-shot both by sample and by family: none of their binaries or family labels appear in training, and they may differ in implementation details (e.g., encryption scheme, threading strategy, and file traversal logic). This setting tests whether SHIELD learns robust filesystem-level signatures of ransomware activity rather than overfitting to family-specific artifacts. We report both offline classification accuracy on these unseen strains and real-time mitigation efficacy (actions-to-halt and pre-halt corruption) in Section~\ref{sec:results}.

    \subsection{Experimental Methodology}
        \label{sec:methodology}
        
        \subsubsection{Host Setup}
        All experiments are executed in a controlled sandbox environment. The sandboxed host and the trusted-side acquisition modules run on a single physical machine equipped with 64\,GB RAM and 1\,TB of storage. The sandbox is provisioned with 16\,GB RAM and a 32\,GB OS disk, running Windows~10. It accesses the disk-under-test (virtual image or FPGA-attached physical drive) as a standard removable network block device. Executables are preloaded inside the sandbox, and we use system snapshots to restore the environment to a clean state between runs, ensuring repeatability and preventing cross-test contamination.
        
        \subsubsection{Data Collection and Labeling}
        To generate datasets under the test architecture, each workload (benign application or ransomware sample) executes for a fixed duration of six minutes while SHIELD records per-action telemetry. Each run follows a consistent workflow where we: (1) clone a baseline disk image, (2) start SHIELD and the trusted-side transport harness, (3) attach the sandbox to the disk-under-test, (4) execute the workload, (5) terminate the run and restore the sandbox from snapshot, and (6) discard the cloned disk image. This procedure ensures that every execution begins from a comparable filesystem state and that observed differences primarily reflect workload behavior rather than residual artifacts.
        Across the ten benign and ten ransomware workloads (Table~\ref{tab:software_lists}), we collect three runs per program, one for each test disk, yielding 30 benign runs and 30 malicious runs for training and in-distribution testing. Windowing and label assignment follow Section~\ref{sec:window_labeling}; sample counts produced by each window configuration are reported in Section~\ref{sec:results_data}.

    \subsection{Evaluation Protocol}
    \label{sec:eval_protocol}
        We evaluate SHIELD along four dimensions that collectively measure (i) discriminative power of the acquired metrics, (ii) alignment with a storage-side (hardware-feasible) deployment, (iii) robustness to emerging threats, and (iv) real-time mitigation effectiveness. We first summarize the total number of recorded actions and the resulting windowed samples for each $(W,O)$ configuration in Section~\ref{sec:results_data}. Using these datasets and deployments, we report the following evaluations:

        \begin{itemize}[noitemsep,nolistsep,leftmargin=*]
            \item \textbf{In-distribution classification:}
            We evaluate six classifiers for both binary (benign vs.\ malicious) and multiclass (ransomware family) performance using the training/test split in Section~\ref{sec:model_training}. Models are evaluated across the window/overlap sweep and the hyperparameter grid (Appendix~\ref{sec:perf_table_full}, Figure~\ref{fig:heatmaps}). We report accuracy, precision, recall, and F1, highlighting (i) binary detection performance in Section~\ref{sec:malicious_benign} and (ii) multiclass family identification in Section~\ref{sec:multiclass_strain}. These experiments establish whether filesystem-aware metrics can reliably separate ransomware from benign high-I/O workloads and capture strain-specific patterns.
            
            \item \textbf{Hardware-only ablation:}
            To assess feasibility under a controller-integrated deployment, we remove transport/interface-derived metrics and retrain models using only filesystem-level features obtainable below the storage interface. We report accuracy and F1 for this hardware-only feature set in Section~\ref{sec:hardware_only}. This ablation isolates the contribution of filesystem-semantic signals and validates that detection does not depend on the PoC transport harness. This ablation further serves to identify efficacy boosts when paired with other defenses.
       
            \item \textbf{Zero-shot generalization to unseen families:}
            We test whether the learned boundary captures general ransomware behavior rather than overfitting to known families by evaluating the best-performing binary model on unseen ransomware families excluded from training (zero-shot by sample and family). We report offline window-level accuracy (and associated metrics) on these unseen strains in Section~\ref{sec:zero_shot}, quantifying generalization to emerging threats.

            \item \textbf{Real-time mitigation efficacy:}
            We deploy the selected binary model in closed-loop mode and measure operational outcomes: (i) detection latency in actions (windows/actions until enforcement triggers) and (ii) the extent of pre-halt corruption. Corruption is measured at both the filesystem level (e.g., number/percentage of affected files) and the storage level (e.g., bytes overwritten) following Section~\ref{sec:realtime}. These results quantify how quickly SHIELD reacts and how tightly it bounds damage under active ransomware execution.
            
            \item \textbf{False positives under benign workloads:}
            To characterize operational safety, we report (i) false-positive rates on benign windowed samples and (ii) the maximum number of consecutive malicious decisions relative to the enforcement threshold (Section~\ref{sec:false_pos}). These tests exist beyond the previously selected ten benign programs used for training/testing, as we additionally run a small exploratory check on five extra benign applications to address generalization to more benign programs; this limited probe is intended only to sanity-check generalization, with broader benign coverage left to future work.

        \end{itemize}

\section{Results \& Evaluation}
    \label{sec:results}
    
    \subsection{Dataset \& Sample Composition} \label{sec:results_data}
\begin{figure}[t]
    \centering
    \begin{minipage}{0.45\linewidth}
        \centering
        \captionof{table}{Summary of Action Window/Overlap Configurations and Subsequent Sample Counts}
        \label{tab:window_summary}
        \resizebox{\linewidth}{!}{
            \setlength\extrarowheight{-3pt}
            \begin{tabular}{rcrrr}
                \toprule
                \textbf{Actions} & \textbf{Overlap} & \textbf{Samples} & \textbf{No. Benign} & \textbf{No. Malicious} \\
                \cmidrule(lr){1-2} \cmidrule(lr){3-5}
                1   & -   & 498{,}230 & 237{,}575 & 260{,}655 \\
                2   & 0   & 249{,}131 & 118{,}794 & 130{,}337 \\
                5   & 0   & 99{,}671  & 47{,}527  & 52{,}144  \\ \cmidrule(lr){1-2} \cmidrule(lr){3-5}
                    & 0   & 49{,}851  & 23{,}771  & 26{,}080  \\
                10  & 2   & 249{,}131 & 118{,}794 & 130{,}337 \\
                    & 5   & 99{,}671  & 47{,}527  & 52{,}144  \\ \cmidrule(lr){1-2} \cmidrule(lr){3-5}
                    & 0   & 24{,}943  & 11{,}893  & 13{,}050  \\
                20  & 4   & 124{,}581 & 59{,}404  & 65{,}177  \\
                    & 10  & 49{,}851  & 23{,}771  & 26{,}080  \\ \cmidrule(lr){1-2} \cmidrule(lr){3-5}
                    & 0   & 16{,}637  & 7{,}933   & 8{,}704   \\
                30  & 5   & 99{,}671  & 47{,}527  & 52{,}144  \\
                    & 15  & 33{,}241  & 15{,}851  & 17{,}390  \\ \cmidrule(lr){1-2} \cmidrule(lr){3-5}
                    & 0   & 9{,}995   & 4{,}767   & 5{,}228   \\
                50  & 10  & 49{,}851  & 23{,}771  & 26{,}080  \\
                    & 25  & 19{,}956  & 9{,}517   & 10{,}439  \\ \cmidrule(lr){1-2} \cmidrule(lr){3-5}
                    & 0   & 5{,}016   & 2{,}391   & 2{,}625   \\
                100 & 20  & 24{,}943  & 11{,}893  & 13{,}050  \\
                    & 50  & 9{,}995   & 4{,}767   & 5{,}228   \\
                \bottomrule
            \end{tabular}
        }
    \end{minipage}\hfill
    \begin{minipage}{0.53\linewidth}
        \centering
        \includegraphics[width=\linewidth]{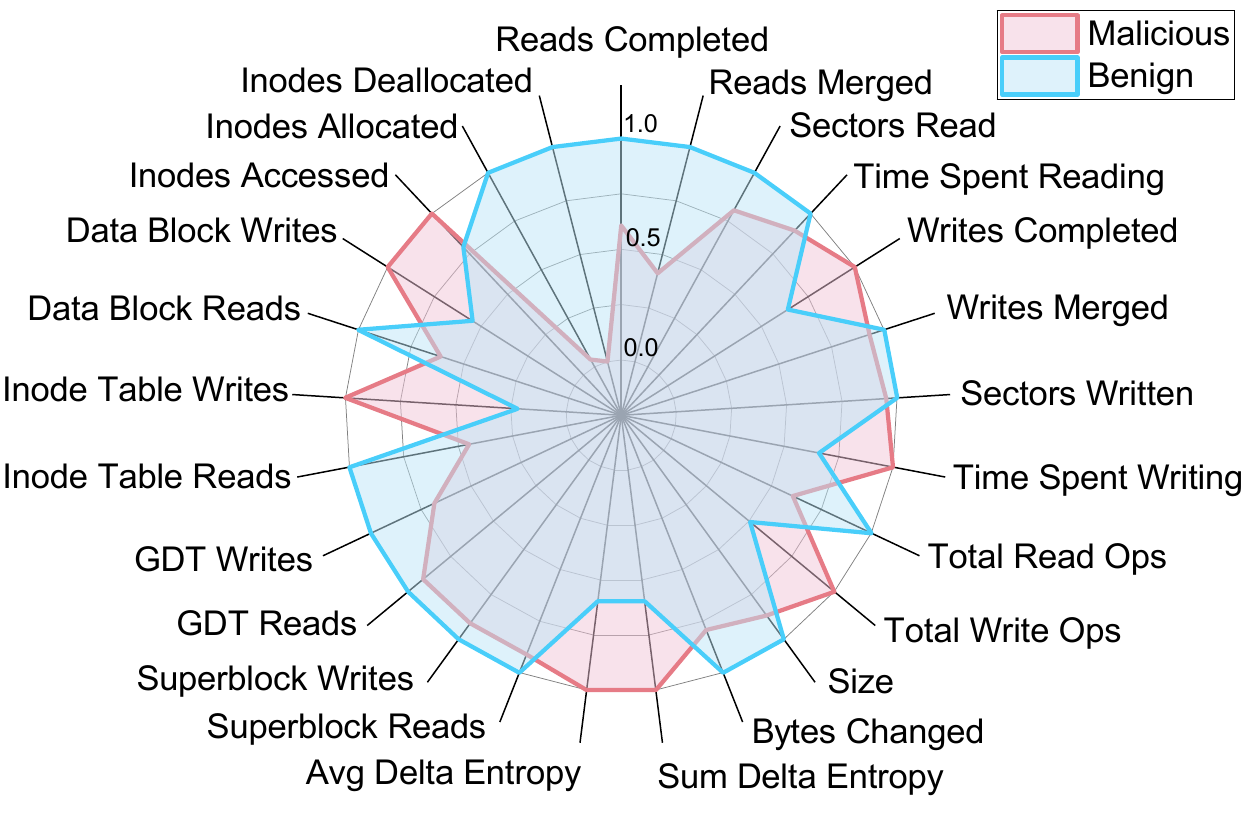}
        \caption{Normalized mean feature values for malicious and benign samples}
        \label{fig:mean_feat}
    \end{minipage}
\end{figure}

        Across 60 executions (30 benign and 30 ransomware), SHIELD recorded 498{,}230 labeled disk actions (one action = one read or write request). These actions comprise of 237{,}575 benign and 260{,}655 malicious samples. To construct model learning inputs, we aggregate the these samples into action-windows of size $W$ with overlap $O$, where consecutive windows begin $W-O$ actions apart (Section~\ref{sec:window_labeling}). We summarize the total samples and window configurations in Table~\ref{tab:window_summary}. 

        Figure~\ref{fig:mean_feat} shows normalized mean feature values for benign versus malicious samples, providing intuition about which signals shift under ransomware activity. Across the dataset, ransomware windows tend to exhibit higher write intensity and larger entropy shifts, alongside elevated counts of \texttt{inodes\_accessed} and data-block writes, which is consistent with multi-file traversal and repeated overwrite patterns. In contrast, inode allocation/deallocation events are comparatively rare in ransomware runs when compared to benign software, reflecting that most samples mutate existing files rather than creating many new ones. Note that these mean shifts are not sufficient on their own to characterize ransomware behavior as many features co-vary (e.g., read volume, write volume, and inode accesses), motivating the multivariate classifiers evaluated in the next subsection rather than assuming feature independence.

    \subsection{In-Distribution Classification Performance} \label{sec:model_accs}
\begin{figure*}[h]
            \centering
            \includegraphics[width=1\textwidth]{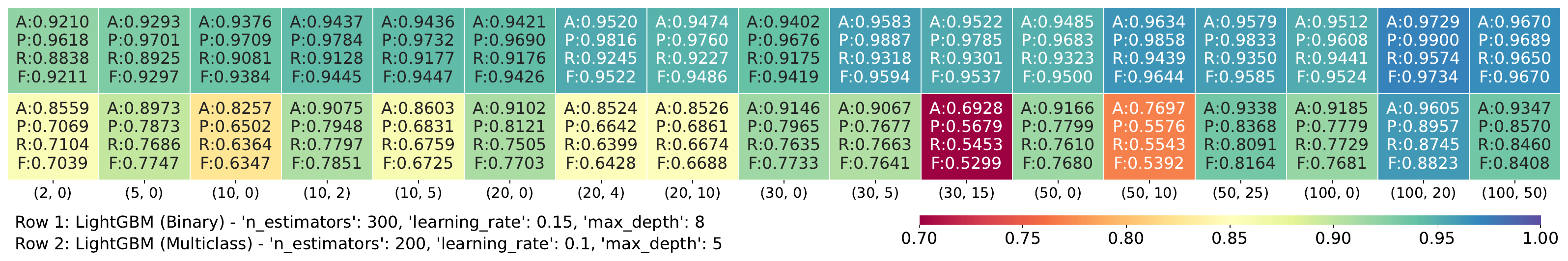}
            \caption{Accuracy-based gradient heatmap for the best binary (top row) and multiclass (bottom row) classifier, showing accuracy (A), precision (P), recall (R), and F1 (F) across window/overlap settings (e.g., $(20\,, 10)$ = 20 actions with 10-action overlap).}
            \label{fig:heatmap_top}
            \vspace{-1em}
        \end{figure*}

        We evaluate SHIELD's windowed feature vectors on in-distribution workloads using six standard classifiers (LightGBM, Random Forest, SVM, KNN, Naive Bayes, and AdaBoost) across the window/overlap configurations in Table~\ref{tab:window_summary}. We focus on the best-performing binary detector and the best-performing multiclass family classifier identified from this sweep. Figure~\ref{fig:heatmap_top} summarizes the performance of these two top models across window/overlap settings; each cell reports accuracy (A), precision (P), recall (R), and F1 (F), with cell color indicating accuracy. The complete performance matrix over all models and configurations is provided in Appendix~\ref{sec:perf_table_full} (Figure~\ref{fig:heatmaps}).

        \subsubsection{Binary Detection (Benign vs Malicious)} \label{sec:malicious_benign}
    
            Across all tested configurations, the best in-distribution binary detector is a LightGBM model (\texttt{n\_estimators}=300, \texttt{learning\_rate}=0.15, \texttt{max\_depth}=8). Peak performance is observed at a \win{100}{20} window, achieving $0.9729$ accuracy and $0.9734$ F1 (precision $=0.9900$, recall $=0.9574$), as summarized in Table~\ref{tab:feature_classification_importance_table} and reflected in Figure~\ref{fig:heatmap_top}. Performance remains consistently strong across moderate-to-large windows, indicating that the acquired filesystem-aware metrics provide stable separation between benign high-I/O activity and ransomware behavior under the training distribution.
                
            In order interpret the main features driving classification, Table~\ref{tab:feature_classification_importance_table} reports split-based feature importance for the best-performing model. The highest-ranked features include \texttt{time\_spent\_reading} and \texttt{size}, along with filesystem-centric indicators such as \texttt{inodes\_accessed} and \texttt{bytes\_changed}. This ranking is consistent with ransomware's access pattern of enumerating many files (inode lookups) followed by repeated overwrite operations that alter block contents and entropy. 
            Finally, because our target deployment is storage-side, we explicitly evaluate whether binary detection remains effective when restricting the feature set to hardware-observable filesystem metrics only. We report this hardware-only ablation and its performance impact in Section~\ref{sec:hardware_only}.

    \subsubsection{Multiclass Family Classification}
    \label{sec:multiclass_strain}
        
        For ransomware family identification, the best multiclass model is LightGBM (\texttt{n\_estimators}=200, \texttt{learning\_rate}=0.1, \texttt{max\_depth}=5) under the \win{100}{20} configuration, achieving $0.9605$ accuracy with a lower F1 of $\approx 0.88$. The F1 gap indicates non-trivial confusion among families whose disk-level behaviors overlap (Figure~\ref{fig:heatmap_top}).
        
        The most discriminative signals for family separation are read- and traversal-oriented features, which reflect how strains enumerate and stage files before encryption. Differences in inode lookups, read volume, and request-size patterns (e.g., few large reads vs.\ many small reads) distinguish broad directory scans from selective targeting and throttled access. In contrast, write-side signals and entropy shifts are more consistent across families once encryption begins, though block-level entropy changes and metadata/data-write activity still add separation for more aggressive strains. Overall, multiclass accuracy is driven primarily by pre-encryption access strategy, whereas binary detection benefits from the shared overwrite signature common to ransomware.

\begin{figure}[t]
    \centering
    \begin{minipage}{0.44\linewidth}
        \centering
        \captionof{table}{Best Model Feature Importance using Full and HW-Only Feature Set}
        \label{tab:feature_classification_importance_table}
        \resizebox{\linewidth}{!}{%
            \setlength\extrarowheight{-3pt}
            \begin{tabular}{l c c}
                \toprule
                 & \textbf{Full Feature Set} & \textbf{HW-Only Feature Set} \\
                \cmidrule(lr){2-2} \cmidrule(lr){3-3}
                \textbf{Model} & LightGBM & LightGBM\\
                \textbf{Window} & \texttt{100/20} (24{,}943 Samples) & \texttt{100/20} (24{,}943 Samples) \\
                \textbf{Accuracy} & $0.9729$ & $0.9597$ \\
                \textbf{Precision} & $0.9900$ & $0.9745$ \\
                \textbf{Recall} & $0.9574$ & $0.9469$ \\
                \textbf{F1 Score} & $0.9734$ & $0.9605$ \\
                \cmidrule(lr){2-2} \cmidrule(lr){3-3}
                \makecell[c]{\textbf{Top 10} \\ \textbf{Features}}
                  & \makecell[l]{%
                    ($1942$) \texttt{time\_spent\_reading}\\
                    ($794 $) \texttt{size}\\
                    ($765 $) \texttt{sectors\_read}\\
                    ($671 $) \texttt{inodes\_accessed}\\
                    ($566 $) \texttt{reads\_merged}\\
                    ($500 $) \texttt{bytes\_changed}\\
                    ($425 $) \texttt{sum\_delta\_entropy}\\
                    ($419 $) \texttt{writes\_merged}\\
                    ($402 $) \texttt{time\_spent\_writing}\\
                    ($396 $) \texttt{reads\_completed}\\
                  }
                  & \makecell[l]{%
                    ($2056$) \texttt{size}\\
                    ($1568$) \texttt{inodes\_accessed}\\
                    ($1159$) \texttt{bytes\_changed}\\
                    ($816 $) \texttt{sum\_delta\_entropy}\\
                    ($778 $) \texttt{total\_read\_ops}\\
                    ($519 $) \texttt{total\_write\_ops}\\
                    ($508 $) \texttt{avg\_delta\_entropy}\\
                    ($375 $) \texttt{data\_block\_writes}\\
                    ($365 $) \texttt{inode\_table\_writes}\\
                    ($350 $) \texttt{data\_block\_reads}\\
                  } \\
                \bottomrule
            \end{tabular}%
        }
    \end{minipage}\hfill
    \begin{minipage}{0.52\linewidth}
        \centering
        \captionof{table}{Unseen Ransomware Samples: Accuracy under Three Window/Overlap Configurations} \vspace{3px}
        \label{tab:unseen_strains}
        \resizebox{0.9\linewidth}{!}{%
            \setlength\extrarowheight{-3pt}
            \begin{tabular}{lrrrr}
                \toprule
                \textbf{Unseen Strain} & \textbf{Samples} & $\mathbf{(2,0)}$ & $\mathbf{(20,10)}$ & $\mathbf{(100,20)}$ \\
                \cmidrule(r{0.2em}){1-2} \cmidrule(l{0.5em}){3-5}
                Akira       & 3{,}649  & $0.8307$ & $0.8986$ & $0.9891$ \\
                Phobos      & 7{,}253  & $0.8260$ & $0.8678$ & $0.9835$ \\
                Inc         & 6{,}141  & $0.7867$ & $0.7772$ & $0.8669$ \\
                DragonForce & 3{,}002  & $0.7695$ & $0.8472$ & $0.9868$ \\
                Trigona     & 16{,}064 & $0.6866$ & $0.7810$ & $0.9527$ \\
                HelloDown   & 11{,}950 & $0.7252$ & $0.7674$ & $0.8763$ \\
                Expiro      & 789      & $0.4633$ & $0.5949$ & $0.9500$ \\
                LokiLocker  & 4{,}472  & $0.8444$ & $0.9129$ & $0.9598$ \\
                Lynx        & 6{,}804  & $0.8398$ & $0.8590$ & $0.9150$ \\
                CryLock     & 9{,}560  & $0.8699$ & $0.9007$ & $0.9790$ \\
                \cmidrule(l{0.5em}){3-5}
                \multicolumn{2}{c}{\textbf{Average}} & $\mathbf{0.7642}$ & $\mathbf{0.8207}$ & $\mathbf{0.9459}$ \\
                \bottomrule
            \end{tabular}%
        }
    \end{minipage}

\end{figure}
    
    \subsection{Hardware-Only Feasibility}
    \label{sec:hardware_only} 
        
        A central objective of SHIELD is to enable detection from a storage-side vantage point, where the host OS (and any host-visible telemetry) is untrusted. To that end, we retrain the best-performing binary classifier using a hardware-only feature set which removes all transport-derived metrics, leaving only filesystem-derived signals obtainable underneath the storage boundary. Table~\ref{tab:feature_classification_importance_table} shows that accuracy remains high under this restriction. Using the same LightGBM architecture and the same \win{100}{20} window configuration, the full feature set (transport plus hardware) achieves $0.9729$ accuracy (F1=$0.9734$), while the hardware-only model attains $0.9597$ accuracy (F1=$0.9605$). The modest reduction confirms that the classifier does not rely on host-side or transport-side artifacts; instead, the dominant discriminative power comes from deep filesystem features that a controller-integrated implementation can observe and log directly.
        
        The shift in top-ranked features further supports this conclusion. Under the full feature set, several high-importance predictors are interface-centric (e.g., \texttt{time\_spent\_reading}, \texttt{sectors\_read}, and request merge counters), reflecting how the PoC transport layer co-varies with workload intensity. Once these are removed, the hardware-only model elevates filesystem-semantic indicators that more directly reflect ransomware behavior: \texttt{inodes\_accessed} and \texttt{bytes\_changed} rise to the top, capturing large-scale file traversal and rapid overwrite activity; entropy-based features (\texttt{sum\_delta\_entropy}, \texttt{avg\_delta\_entropy}) remain highly ranked, reflecting the transition from low-entropy plaintext to high-entropy ciphertext; and block-level access counts (\texttt{data\_block\_writes}, \texttt{data\_block\_reads}) and metadata mutation signals (\texttt{inode\_table\_writes}) appear among the top predictors. Notably, these features correspond to operations that ransomware must perform to succeed, making them difficult to suppress without substantially degrading the attack---we identify how these semantics relate to ransomware behavior in Section~\ref{sec:fs_importance_discuss}.
        Overall, this ablation provides evidence that in-controller deployment is feasible as performance is retained using only storage-side telemetry. The most influential predictors in this case still align with expected semantics of multi-file encryption rather than prototype or test harness artifacts.

    \subsection{Zero-Shot Detection Ability} 
        To evaluate generalization beyond the training distribution, we perform zero-shot testing on ten ransomware families unrelated to the original training set (zero-shot by both sample and family). We report two complementary evaluations: (i) an `offline' analysis that measures classification accuracy on the complete logs from each unseen strain, and (ii) an `online' closed-loop test in which the same pre-trained model is deployed and allowed to halt writes after sustained malicious decisions, thereby allowing for feasibility assessment and measurement of practical damage bounds.

\newcommand*\rotl[1]{\hbox to1em{\hss\rotatebox[origin=br]{-40}{#1}}}

\newcommand{\qmidrule}{
    \cmidrule(lr){1-2} \cmidrule(lr){3-5} \cmidrule(lr){6-8} \cmidrule(lr){9-9} \cmidrule(lr){10-11} \cmidrule(lr){12-13} \cmidrule(lr){14-15}
    }

\begin{table*}[ht]
\centering
\caption{Detection and Data Corrupted (in Bytes) for Unseen Ransomware Strains across Different  Windows}
\label{tab:realtime_unseen}
\resizebox{\linewidth}{!}{%
\setlength\extrarowheight{-3pt}
\begin{tabular}{ccrrrrrrrrrrrrr}
\textbf{Unseen Strain} & \textbf{Window} &
\rotl{\textbf{Decisions-to-Detect (DTD)}}   & 
\rotl{\textbf{Decisions-Bengign (DB)}}         & 
\rotl{\textbf{Decisions-Maliciouis (DM)}} &
\rotl{\textbf{Actions-to-Detect (ATD)}}        & 
\rotl{\textbf{Total Reads}}      & 
\rotl{\textbf{Total Writes}}     & 
\rotl{\textbf{Time-to-Detect (TTD)}, ms}   &
\rotl{\textbf{Files Affected (FA)}}         & 
\rotl{\textbf{(\%FA)}} \rotl{\textbf{Percent Files Affected}}      &
\rotl{\textbf{(MA-OS)}, bytes} \rotl{\textbf{Memory Affected in OS }}      & 
\rotl{\textbf{(\%MA-OS)}} \rotl{\textbf{Percent Mem. Aff. in OS }}    &
\rotl{\textbf{(MA-HW)}, bytes} \rotl{\textbf{Memory Affected in HW }}      & 
\rotl{\textbf{(\%MA-HW)}} \rotl{\textbf{Percent Mem. Aff. in HW }} \\
\qmidrule

 & 2/0      & 6  & 2  & 4 & 12  & 11  & 1  & 1950 & 2  & 0.0091\% & 244{,}531         & 0.0024\% & 2{,}715   & $<$0.0001\% \\    
{{Akira}}  & 20/10    & 7  & 2 & 5 & 140 & 85    & 55 & 6808 & 56 & 0.2550\% & 322{,}458         & 0.0031\% & 943{,}748         & 0.0092\% \\   
 & 100/20   & 4  & 0 & 4 & 400 & 186  & 214 & 14515 & 71  & 0.3233\% & 27{,}577{,}548  & 0.2677\% &  2{,}940{,}604  & 0.0285\% \\
\qmidrule

 & 2/0      & 29 & 22 & 7 & 58  & 46  & 12 & 5604 & 25 & 0.1138\% & 7{,}234{,}560     & 0.0013\% & 130{,}944 & 0.0012\% \\
{{Phobos}}  & 20/10    & 9  & 5 & 4 & 180 & 117   & 63 & 6252 & 51 & 0.2322\% & 20{,}552{,}089    & 0.1995\% & 468{,}726         & 0.0046\% \\   
 & 100/20   & 5  & 1 & 4 & 500 & 265  & 235 & 15233 & 118 & 0.5373\% & 54{,}735{,}667  & 0.5314\% &  3{,}963{,}504  & 0.0385\% \\
\qmidrule

 & 2/0      & 33 & 24 & 9 & 66  & 51  & 15 & 5288 & 2  & 0.0091\% & 876{,}339         & 0.0085\% & 380       & $<$0.0001\%  \\    
{{Inc}} & 20/10    & 11 & 7 & 4 & 220 & 160   & 60 & 8436 & 19 & 0.0865\% & 6{,}291{,}456     & 0.0611\% & 1{,}212{,}255     & 0.0118\% \\
 & 100/20   & 5  & 1 & 4 & 500 & 271  & 229 & 16258 & 70  & 0.3188\% & 81{,}788{,}928  & 0.7940\% &  4{,}815{,}220  & 0.0467\% \\
\qmidrule

 & 2/0      & 39 & 31 & 8 & 78  & 70  & 8  & 5461 & 46 & 0.2095\% & 19{,}608{,}371    & 0.1904\% & 1{,}789   & $<$0.0001\%  \\    
{{DragonForce}} & 20/10    & 9  & 4 & 5 & 180 & 128   & 52 & 7604 & 61 & 0.2778\% & 26{,}528{,}972    & 0.2575\% & 729{,}665         & 0.0071\% \\
 & 100/20   & 4  & 0 & 4 & 400 & 211  & 189 & 13527 & 93  & 0.4235\% & 48{,}234{,}496  & 0.4683\% &  3{,}935{,}358  & 0.0382\% \\
\qmidrule

 & 2/0      & 74 & 69 & 5 & 148 & 134 & 14 & 5053 & 85 & 0.3871\% & 45{,}508{,}198    & 0.4418\% & 441    & $<$0.0001\%  \\  
{{Trigona}} & 20/10    & 7  & 3 & 4 & 140 & 58    & 82 & 5252 & 80 & 0.3643\% & 18{,}979{,}225    & 0.1843\% & 946{,}463         & 0.0092\% \\
 & 100/20   & 5  & 1 & 4 & 500 & 198  & 302 & 4101  & 125 & 0.5692\% & 66{,}689{,}433  & 0.6474\% &   125{,}207     & 0.0012\% \\
\qmidrule

 & 2/0      & 59 & 54 & 5 & 118 & 103 & 15 & 5084 & 81 & 0.3689\% & 40{,}999{,}321    & 0.3980\% & 430    & $<$0.0001\%  \\    
{{HelloDown}} & 20/10    & 10 & 6 & 4 & 200 & 136   & 64 & 6384 & 100& 0.4554\% & 49{,}492{,}787    & 0.4805\% & 212{,}398         & 0.0021\% \\ 
 & 100/20   & 5  & 1 & 4 & 500 & 218  & 282 & 11122 & 138 & 0.6284\% & 72{,}771{,}174  & 0.7065\% &   955{,}985     & 0.0093\% \\
\qmidrule

 & 2/0      & 39 & 30 & 9 & 78  & 62  & 16 & 3796 & 20 & 0.0911\% & 4{,}460{,}032     & 0.0433\% & 19{,}068  & 0.0002\% \\   
{{Expiro}} & 20/10    & 8  & 4 & 4 & 160 & 108   & 52 & 6429 & 74 & 0.3370\% & 40{,}265{,}318    & 0.3909\% & 256{,}982         & 0.0025\% \\
 & 100/20   & 5  & 1 & 4 & 500 & 202  & 298 & 13215 & 60  & 0.2732\% & 26{,}214{,}400  & 0.2545\% & 11{,}007{,}997  & 0.1069\% \\
\qmidrule

 & 2/0      & 28 & 21 & 7 & 56  & 34  & 22 & 5430 & 13 & 0.0592\% & 4{,}928{,}307     & 0.0478\% & 3{,}990   & $<$0.0001\%  \\     
{{LokiLocker}} & 20/10    & 5  & 1 & 4 & 100 & 59    & 41 & 5161 & 19 & 0.0865\% & 6{,}291{,}456     & 0.0611\% & 730{,}138         & 0.0071\% \\
 & 100/20   & 4  & 0 & 4 & 400 & 204  & 196 & 19010 & 95  & 0.4326\% & 45{,}613{,}056  & 0.4428\% &  3{,}666{,}591  & 0.0356\% \\
\qmidrule

 & 2/0      & 60 & 52 & 8 & 120 & 110 & 10 & 5362 & 1  & 0.0046\% & 208{,}691         & 0.0020\% & 232    & $<$0.0001\%  \\     
{{Lynx}} & 20/10    & 12 & 8 & 4 & 240 & 173   & 67 & 6945 & 57 & 0.2596\% & 321{,}536         & 0.0031\% & 728{,}499         & 0.0071\% \\
 & 100/20   & 5  & 1 & 4 & 500 & 261  & 239 & 4959  & 225 & 1.0246\% & 92{,}694{,}118  & 0.8999\% &  1{,}206{,}425  & 0.0117\% \\
\qmidrule

 & 2/0      & 50 & 41 & 9 & 100 & 98  & 2  & 4616 & 42 & 0.1913\% & 11{,}010{,}048    & 0.1069\% & 1{,}882   & $<$0.0001\%  \\      
{{CryLock}} & 20/10    & 9  & 5 & 4 & 180 & 112   & 68 & 7340 & 47 & 0.2140\% & 1{,}541{,}4067    & 0.1496\% & 60{,}293          & 0.0006\% \\ 
 & 100/20   & 6  & 2 & 4 & 600 & 307  & 293 & 15362 & 146 & 0.6648\% & 73{,}610{,}035  & 0.7146\% &  2{,}578{,}270  & 0.0250\% \\
\cmidrule(lr){1-15}
 & 2/0      & 41.7 & 34.6 & 7.1 & 83.4 & 71.9 & 11.5 & 4764.4 & 31.7 & 0.1444\% & 1{,}350{,}840 & 0.1242\% & 16{,}187 & 0.0002\% \\   
 {\textbf{Average}} & 20/10    & 8.7 & 4.5 & 4.2 & 174 & 113.6 & 60.4 & 6661.1 & 56.4 & 0.2568\% & 18{,}445{,}936 & 0.179\%1 & 628{,}917 & 0.0061\% \\ 
 & 100/20   & 4.8 & 0.8 & 4 & 480 & 232.3 & 247.7 & 12730.2 & 114.1 & 0.5196\% & 58{,}992{,}886 & 0.5727\% & 3{,}519{,}516 & 0.0342\% \\
\bottomrule

\end{tabular}
} 
\end{table*}

        \subsubsection{Offline Zero-shot Accuracy} \label{sec:zero_shot}
            Table~\ref{tab:unseen_strains} reports the binary detector's accuracy on windowed samples derived from unseen-strain execution traces under three representative window configurations \win{2}{0}, \won{20}{10}, and \won{100}{20}. As expected, larger windows yield higher accuracy due to more behavioral context per decision: the average accuracy increases from $0.7642$ at a \won{2}{0} window to $0.8207$ at \won{20}{10}, and reaches $0.9459$ when \win{100}{20}. This trend is particularly evident for strains whose activity is sporadic or intermittent (e.g., Expiro), where very small windows may capture only partial stages of the workflow. Nevertheless, even with small windows the model identifies malicious behavior early in execution, motivating the online mitigation experiment below, where earlier detection directly translates into reduced data loss.

        \subsubsection{Closed-loop Mitigation on Unseen Strains}
        \label{sec:realtime}
            
\begin{figure}[t]
                \centering
                \begin{minipage}[t]{0.49\linewidth}
                    \centering
                    \includegraphics[width=\linewidth]{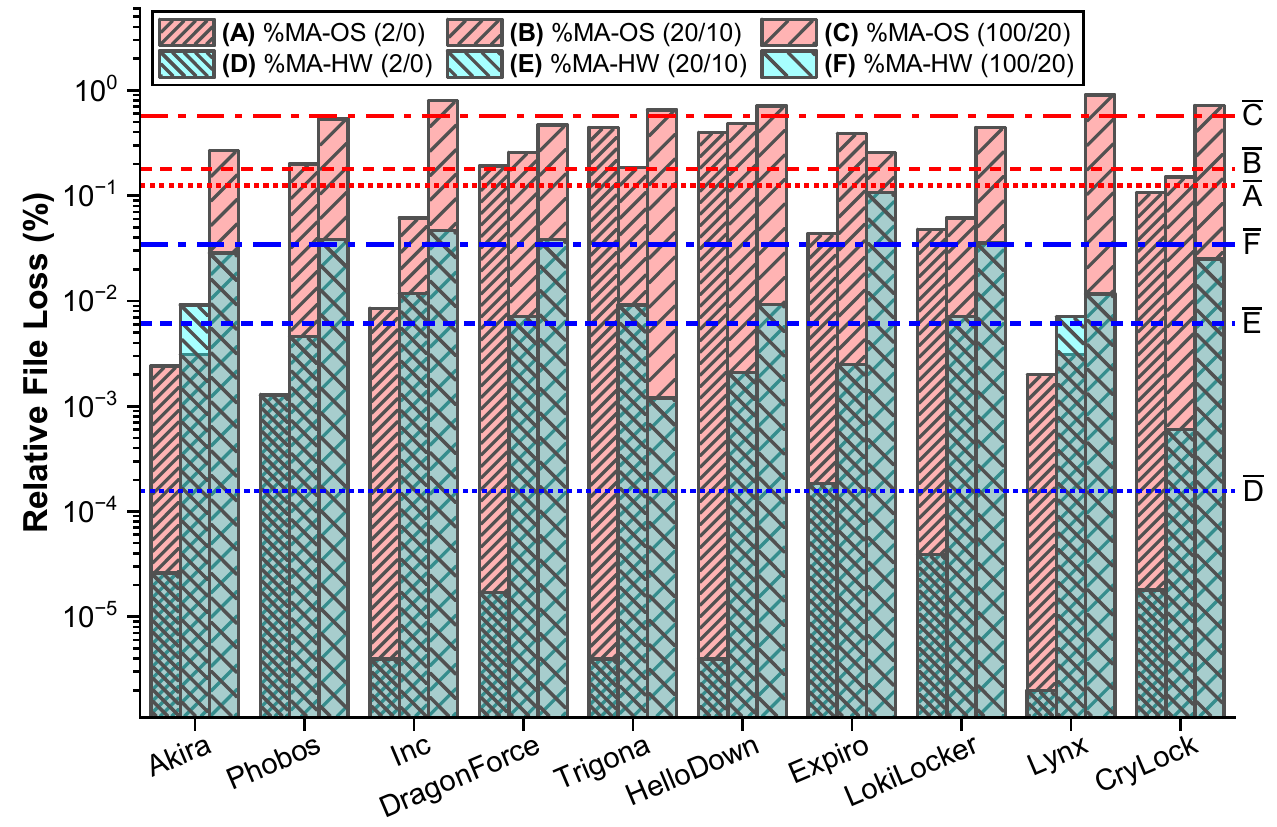} \vspace{-2em}
                    \caption{Comparison of \%MA-OS and \%MA-HW overlapped across unseen strains for three window settings, along with lines depicting mean}
                    \label{fig:unseen_memenc}
                \end{minipage}\hfill
                \begin{minipage}[t]{0.49\linewidth}
                    \centering
                    \includegraphics[width=\linewidth]{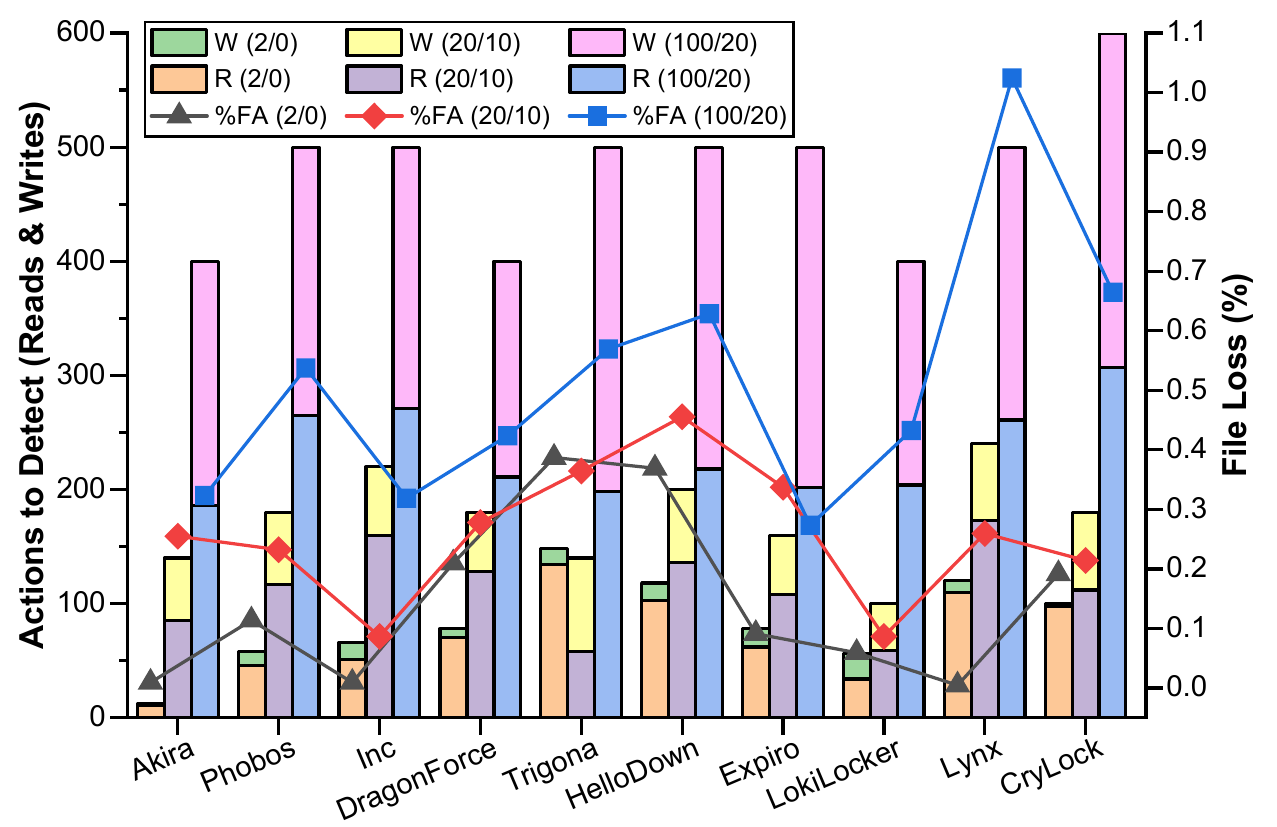} \vspace{-2em}
                    \caption{Actions-to-detect (ATD) made up of reads and writes vs. Relative Files Affected (\%FA) for each unseen strain,
                    comparing three window settings}
                    \label{fig:unseen_atdfile}
                \end{minipage}
            \end{figure}

            We next deploy the same pre-trained LightGBM binary detector in the online SHIELD enforcement loop and execute each unseen strain end-to-end. Unlike the offline analysis, which scores all windowed samples to give an overall accuracy, the online system makes decisions sequentially and halts the disk after only a prefix of the trace. Table~\ref{tab:realtime_unseen} summarizes the resulting detection and damage bounds across three $(W,O)$ configurations. Figures~\ref{fig:unseen_memenc} and ~\ref{fig:unseen_atdfile} visualize detection effort and pre-halt damage across strains and window configurations.

            Across all ten unseen families, SHIELD halts disk writes before widespread corruption, demonstrating that the learned decision boundary transfers to previously unseen ransomware behaviors when driven by deep filesystem telemetry. As in the offline study, the action window controls a fundamental responsiveness--confidence trade-off. Smaller windows (e.g., \won{2}{0}) require more sequential decisions to reach the halt threshold, but they halt after fewer total actions on average, reducing the opportunity for encryption writes to accumulate. This pattern is visible in Figure~\ref{fig:unseen_atdfile}: \win{2}{0} consistently yields the lowest actions-to-detect (ATD) across strains and correspondingly lower relative files affected (\%FA). In contrast, larger windows (e.g., \win{100}{20}) reduce the number of decisions needed to trigger a halt, but they aggregate more actions per decision, which increases ATD and typically raises \%FA for strains with faster write phases (e.g., Lynx). The intermediate setting (e.g., \win{20}{10}) provides a middle operating point.

            Beyond file counts, Figure~\ref{fig:unseen_memenc} highlights a consistent discrepancy between OS-level and device-level damage accounting. Across all three window settings, the percentage of bytes the OS deems corrupted (\%MA-OS) exceeds the percentage of bytes actually overwritten on disk (\%MA-HW), often by orders of magnitude. This gap occurs because ransomware frequently touches files incrementally while the OS accounts for corruption at file granularity. By measuring changes at the block level, SHIELD can halt during these early write phases before full-file rewrites complete, so even when multiple files are flagged by the OS, the true on-disk overwrite volume remains small. Moreover, this effect persists across window sizes: larger windows increase the time-to-intervention, shifting both \%MA-OS and \%MA-HW upward, but \%MA-HW remains consistently far lower, indicating that the framework is intercepting ransomware during partial encryption rather than after bulk data replacement. 
            Figure~\ref{fig:unseen_atdfile} further interprets strain-to-strain reads, writes and action counts against relative file loss and window sizes, indicating variability in detection effort. Strains that exhibit heavier pre-encryption scanning inflate read-side activity (higher read components of ATD), while strains that transition quickly to encryption manifest as earlier write bursts. Moreover, smaller windows yield samples with more reads than writes, likely a contributing factor into why smaller windows exhibit lower offline accuracy.

    \subsection{Summary \& Key Results}

        \begin{enumerate}[noitemsep,nolistsep,leftmargin=*]
            \item \textbf{Ransomware induces distinct filesystem-level signatures:} Across 60 runs, benignware and ransomware exhibit consistently different access and mutation patterns at the inode, metadata, and data-block layers, yielding separable distributions in the acquired metrics.
            \item \textbf{Detection hinges on multi-feature interactions captured by tree ensembles:} Individual signals (e.g., entropy deltas, inode activity) are useful, but robust ransomware detection requires modeling their joint, non-linear relationships across action windows. Accordingly, independence-based models (e.g., Naive Bayes) underperform, while tree ensembles such as LightGBM and Random Forest consistently achieve the strongest in-distribution results.
            \item \textbf{Deep filesystem features remain effective without OS/transport telemetry:} In the hardware-only ablation, accuracy remains high with the most influential predictors shifting toward storage-observable signals such as \texttt{inodes\_accessed}, \texttt{bytes\_changed}, metadata writes, and block-level entropy change. Notably, access size and inode activity rank above entropy alone, suggesting that traversal and overwrite dynamics can be more discriminative than ciphertext indicators commonly emphasized in prior work.
            \item \textbf{Zero-shot detection enables practical containment, with a tunable accuracy--damage trade-off:} The model generalizes to unseen ransomware families, and closed-loop halting limits corruption before widespread encryption. Smaller action windows react sooner (fewer actions before halt), while larger windows improve confidence and offline accuracy at the cost of allowing more pre-halt writes---providing a clear design knob for deployment.
        \end{enumerate}

\section{Discussion} \label{sec:discussion}

    \subsection{Framework Utility \& Analysis}

        \subsubsection{Discerning Malicious Behavior} 
            Deep filesystem metrics provide a strong and consistent signal for distinguishing ransomware from benign activity. Across the evaluated window configurations, tree-based classifiers trained on our feature set achieve high in-distribution performance for binary detection (up to $\approx$97\% accuracy/F1) and also support multiclass family identification, indicating that filesystem-observable access and mutation patterns capture both malice and behavioral temporality. This signal transfers beyond the training distribution: in zero-shot tests on previously unseen families, the same pre-trained detector maintains high accuracy and enables reliable early detection in closed loop, demonstrating that the learned boundary generalizes to novel ransomware behaviors.
        
        \subsubsection{Hardware-only Feasibility}
            The discriminative power is preserved when restricting the model to features observable below the storage interface. When we omit transport- and storage-server-derived counters (e.g., \texttt{time\_spent\_reading}, \texttt{reads\_merged}, \texttt{writes\_merged}) and retrain using only filesystem-level features in the HW-only ablation (See Section~\ref{sec:hardware_only}), performance remains high ($>95\%$ accuracy and F1; Table~\ref{tab:feature_classification_importance_table}). The most influential hardware-visible predictors emphasize access intensity and mutation semantics (\texttt{size}, \texttt{inodes\_accessed}, \texttt{bytes\_changed}, entropy deltas, and data/metadata write activity) and are all directly observable below the host OS. This result supports our controller-integrated deployment model (Figure~\ref{fig:trust_poc}) where the core detection signal does not depend on trusting the host kernel or any on-host instrumentation.
        
        \subsubsection{Tunable Detection vs. Damage Tradeoff}
            Action-based windowing exposes a tunable responsiveness vs. confidence trade-off that matters for enforcement. Larger windows provide more behavioral context per inference and typically improve offline accuracy, but they also delay intervention by accumulating more actions (and therefore potential writes) before a decision can trigger a halt. Conversely, smaller windows reduce the number of actions observed prior to detection and can limit pre-halt corruption, at the cost of requiring more sequential decisions to reach the halt threshold. This trade-off is reflected both in classification performance and in the observed damage metrics under closed-loop mitigation, making window size and overlap an explicit control for balancing detection confidence against containment speed.

    \subsection{Security Implications \& Trust}
        
        \subsubsection{Tamper resistance under host compromise} 
            Because metric acquisition and enforcement occur off-host, a root-level adversary cannot disable the monitor, forge its observations, or retroactively alter logged events. We view these storage-visible filesystem metrics as device-side indicators of compromise that remain observable even under host compromise. This follows the principle that non-intrusive, hardware-side monitoring can serve as a trusted anchor for runtime anomaly detection in embedded systems~\cite{Lu2017TECSAnomaly}; SHIELD applies this principle at the storage layer, where filesystem-semantic telemetry provides the trust anchor rather than processor execution traces. The host is presented only with a standard block-device interface (reads/writes), and no control channel is exposed for reconfiguring detection logic or modifying internal state. An attacker can still attempt denial-of-service against the storage path, but this primarily impedes the ransomware's own ability to perform encryption and does not enable evasion of the on-disk transitions that SHIELD observes.

        \subsubsection{Data access and locality} 
            SHIELD is designed to avoid exporting file contents or metadata beyond what is necessary for detection. In the current prototype, we compute lightweight byte-wise entropy statistics and filesystem-structure mutations, rather than transferring payload data or reconstructing user files. As a result, sensitive data remains local to the deployment environment, aligning with on-premises security and governance requirements.

        \subsubsection{Transport-agnostic trust boundary} \label{sec:trust_discuss}
            The NBD-based path used in our PoC serves as an evaluation harness that provides a convenient, standard block interface for trace collection and closed-loop testing; it is not a required component of the security model. In a target deployment, SHIELD resides below (or within) the storage controller, preserving the same trust boundary without relying on a network transport. The only requirement is visibility into the block I/O stream (operation type, logical location, size, and write payload) and access to the underlying storage media to interpret filesystem structures. This makes the approach compatible with multiple interconnects (e.g., SATA, NVMe, USB, flash controllers) while keeping the enforcement point and telemetry generation on the trusted storage side.

        \subsubsection{System- and speed-independence via action-based decisions} 
            SHIELD aggregates telemetry over a fixed number of disk actions rather than elapsed time, making detection behavior largely invariant to host load, virtualization overhead, controller throughput, or transport latency. This action-window formulation ties inference to the work performed on the disk, not to wall-clock timing (total elapsed time), and so the same acquisition framework and decision policy remains effective across platforms with different performance characteristics. It also limits an attacker's ability to evade by slowing down encryption or inserting delays: throttling changes the time between actions, but not the fact that the same sequence of on-disk mutations must occur to encrypt files, and those mutations remain visible to the storage-side monitor.

    \subsection{Interpreting Filesystem-semantic Signals} \label{sec:fs_importance_discuss}      
        Feature importance indicates that SHIELD's most discriminative signals are filesystem-semantic measures of structures changed on disk, and not simply disk utilization or how busy the system is. In the hardware-only model (Table~\ref{tab:feature_classification_importance_table}), \texttt{size}, \texttt{inodes\_accessed}, and \texttt{bytes\_changed} dominate feature-split importance: sustained large I/Os reflect bulk overwrite pressure, inode-table accesses capture file-by-file mutation patterns that many ransomware families induce via in-place modification rather than frequent create/delete churn, and \texttt{bytes\_changed} directly measures overwrite magnitude (useful for intermittent or throttled encryption, for example in Expiro ransomware, where write rate alone can be low and misleading). \texttt{sum\_delta\_entropy} remains a strong contributor, but largely as corroboration. Entropy shifts help confirm encryption when aligned with inode and overwrite activity, yet are not uniquely identifying because benign workloads (compression, media processing, encrypted containers) can also exhibit high entropy. Overall, the superior performance of multivariate tree-based models suggests that the key signal is the co-occurrence of large writes, inode/metadata mutation, and substantial block-level change, which an off-host filesystem-aware monitor/mitigator can observe reliably even under full host compromise. This result echoes findings in hardware-based malware detection, where ML classifiers trained on tamper-resistant hardware-observable features outperform those relying on host-level software metrics~\cite{Kadiyala2020TECSHPC}; SHIELD extends this principle from processor performance counters to filesystem-semantic storage telemetry.

    \subsection{Practical Deployment}
    
        \subsubsection{Overhead} \label{sec:overhead}
            Our current implementation is a correctness-first software PoC that layers metric parsing, logging, and inference on top of a virtualized I/O path. Consequently, its wall-clock throughput reflects prototype overhead rather than an optimized controller implementation. Table~\ref{tab:nbd_throughput} reports the measured end-to-end throughput of this PoC under three configurations: baseline transport, transport plus metric logging, and transport plus logging and online inference. Writes incur the largest slowdown because our PoC performs per-write work that is deliberately conservative (per-block entropy calculation and filesystem-structure updates), while reads remain largely unchanged. These measurements are interpreted as an upper bound on overhead in our software evaluation harness: they include VM scheduling effects, network-stack overhead, and unoptimized parsing that would  be substantially reduced or fully removed in a storage-controller pipeline.
            
            Moreover, since SHIELD's detection latency is evaluated in disk actions rather than elapsed time (e.g., actions-to-detect in Section~\ref{sec:realtime}), this decouples the security outcome from machine speed, virtualization overhead, or interconnect latency. In a target FPGA/ASIC deployment, the same stages are naturally streaming and parallelizable, as indexed metadata lookups, lightweight block statistics (including approximate entropy), and shallow model inference can be overlapped with DMA and host I/O to approach near-native throughput.            
            
            \begin{table}[ht]
                \centering
                \caption{End-to-end throughput of the software PoC harness as instrumentation increases: baseline transport, transport with metric logging, and transport with logging plus online inference. Measurements include VM and interface overhead, representing an upper latency bound compared to a controller-integrated pipeline.}
                \label{tab:nbd_throughput}
                \resizebox{0.6\linewidth}{!}{%
                \begin{tabular}{lccc}
                    \toprule
                    \textbf{Operation} & \textbf{Transport Only} &  \textbf{+ Logging} & \textbf{ + Logging \& Inference} \\
                    \midrule
                    Read  & 3700\,MB/s & 3600\,MB/s & 3600\,MB/s \\
                    Write & 2900\,MB/s & 1200\,MB/s & 890\,MB/s \\
                    \bottomrule
                \end{tabular}} \vspace{-1em}
            \end{table}

    \subsubsection{Zero-shot Benignware \& False Positive Rate} \label{sec:false_pos}
        To estimate operational robustness, we evaluated the deployed binary detector on five benign applications that were not used during training (Table~\ref{tab:zeroshot_benign}). Using the \texttt{20/10} configuration, the model exhibits low false-positive rates (all $\leq 3.6\%$) and, critically, does not accumulate enough consecutive malicious decisions to trigger enforcement: the maximum observed run produced two sequential malicious classifications, which is below the four-of-five threshold required to halt writes in the deployment enforcement scheme. This suggests that even a modest benign training set can yield a conservative operating point for common high-I/O workloads, while still preserving aggressive response to ransomware (Section~\ref{sec:realtime}). In practice, false positives can be further reduced by broadening benign coverage (more applications and OS-specific behavior), adding deliberate `benign stress' traces (backup, sync, or compression), or calibrating the enforcement threshold to a determined risk tolerance value.

        \begin{table}[h]
        \centering
        \caption{False-positive behavior (\%FP) on unseen benign applications using the deployed \texttt{20/10} model overall false-positive rate, along with the maximum number of sequential malicious decisions (Seq.\ DM).}
        \label{tab:zeroshot_benign}
        \resizebox{0.6\linewidth}{!}{%
        \setlength\extrarowheight{-3pt}
        \begin{tabular}{lccccc}
        \toprule
        \textbf{Application} & \textbf{Actions} & \textbf{Samples} & \textbf{DMs} & \textbf{FP Rate} & \textbf{Seq. DMs} \\
        \midrule
        Code Blocks         & 13{,}360 & 668 & 14 & 2.10\% & 1 \\
        Web Browser         & 8{,}900  & 445 & 12 & 1.80\% & 1 \\
        Word Processor      & 9{,}640  & 482 & 6  & 0.90\% & 1 \\
        Python IDLE         & 9{,}520  & 476 & 24 & 3.59\% & 2 \\
        PDF Editor         & 10{,}240 & 512 & 8  & 1.20\% & 1 \\
        \bottomrule
        \end{tabular}} \vspace{-1em}
        \end{table}

    \subsection{Filesystems \& Backend Modularity}
        Although our evaluation focuses on \texttt{ext4}, SHIELD is designed around a filesystem-agnostic event schema that captures roles and update patterns rather than OS-specific structures: directory traversal, per-file metadata mutations, allocation/extent updates, and data-block writes ( Table~\ref{tab:fs_feats}). This abstraction is what makes the approach portable across filesystems that implement analogous primitives with different on-disk representations (e.g., \texttt{ext4} inodes vs.\ \texttt{NTFS} MFT records vs.\ \texttt{APFS} object records). Concretely, the metric acquisition framework consumes block I/O and maps offsets to semantic regions, producing the same normalized feature vector regardless of backend. To demonstrate feasibility beyond \texttt{ext4}, we implemented an additional \texttt{NTFS} backend using \texttt{NTFS-3G} that replaces the \texttt{ext4} parsing layer with \texttt{NTFS}-equivalent structure parsing and metric collection, while leaving feature aggregation, learning, and enforcement unchanged~\cite{ntfs-3g}. We do not claim accuracy differences across filesystems but the modular backend confirms that SHIELD's detection pipeline is not tied to a single filesystem and can be extended as engineering effort permits.

    \subsection{Limitations}
        Our results should be interpreted in the context of several limitations. First, throughput and wall-clock overhead are measured in a software PoC evaluation harness (VM + transport + unoptimized parsing) and therefore do not reflect an optimized controller datapath; while the design is intentionally streaming and hardware-friendly, we do not report an end-to-end FPGA/ASIC throughput implementation. Second, the experiments and learned models are validated on \texttt{ext4}; while we built an \texttt{NTFS} backend to demonstrate modularity, a full cross-filesystem evaluation is currently out of scope, and we intentionally avoid making claims about which filesystem yields `better' detection. Third, the dataset scope is necessarily bounded, as program selection, fixed run duration, and a finite set of ransomware/benign workloads may not cover all enterprise behaviors (e.g., backups, database access, or mixed workloads)---we scope these for future work. Fourth, our threat model excludes physical access, side-channel attacks, and supply-chain compromise of the storage-side hardware; these are orthogonal but important risks for deployments in untrusted environments. Finally, like any behavior-based detector, SHIELD can be challenged by adaptive evasion strategies (e.g., slow encryption, interleaving benign I/O, or rate-limiting writes); action-based windows and multi-decision enforcement reduce sensitivity to timing and help smooth intermittent behavior, but they cannot eliminate all adversarial trade-offs between stealth and progress---an attacker willing to encrypt extremely slowly may reduce detectability, albeit at the cost of making the attack less practical and more interruptible.

    \subsection{Future Work}
        We see three near-term directions that extend SHIELD while preserving its storage-side trust boundary. First, we plan to integrate metric acquisition and enforcement directly into a storage controller (FPGA/ASIC). The HW-only results (Table~\ref{tab:feature_classification_importance_table}) indicate that high accuracy does not depend on OS- or transport-derived counters, supporting a tamper-resistant on-disk implementation. A next step is to port the C modules and exported model logic to HDL/HLS, where metadata attribution, approximate entropy, and inference can be pipelined and parallelized to eliminate software and virtualization overhead.
        Second, because SHIELD already observes block contents around overwrites (for byte-difference and entropy features), we can add a bounded rolling buffer of the last $n$ modified blocks. If ransomware is detected, this buffer could enable partial rollback of recently overwritten data, reducing residual damage to touched files.
        Finally, we will harden and validate additional filesystem parsers (e.g., \texttt{NTFS}, \texttt{APFS}) behind the same normalized event schema (Table~\ref{tab:fs_feats}). Immediate work includes an end-to-end evaluation of our \texttt{NTFS}-3G backend to confirm semantic metric equivalence under realistic workloads, and exploring \texttt{APFS} support where tooling permits. The goal is portable telemetry with unchanged learning and enforcement, rather than claims that any filesystem is inherently “more detectable.”

\section{Conclusion} \label{sec:conclusion}
    We presented SHIELD, an embedded, storage-side telemetry and enforcement framework for ransomware detection that remains effective under full host compromise. By parsing filesystem structures and aggregating per-action features into action-windowed samples, SHIELD learns discriminative disk-level signatures of ransomware behavior without relying on on-host instrumentation. In our evaluation across multiple ransomware families and benign workloads, the best models achieve high in-distribution accuracy, and a hardware-only ablation using only storage-visible filesystem features retains $>95\%$ accuracy, confirming feasibility for FPGA/ASIC deployment within the storage controller datapath. In closed-loop experiments on previously unseen families, the same pre-trained detector triggers write blocking after only a prefix of execution, limiting pre-halt corruption. Overall, these results show that deep, filesystem-aware telemetry at the storage boundary can provide robust detection and practical containment while preserving a tamper-resistant trust boundary suitable for embedded storage controller integration.

\clearpage\newpage

\bibliographystyle{ACM-Reference-Format}
\bibliography{refs}

\appendix
\newpage

\section{Pseudocode for Metric Acquisition Algorithm} \label{sec:algo}

\begin{algorithm}[h]
\footnotesize
\caption{\textsc{Shield} Flow for Metric Acquisition}\label{alg:acquisition}

\KwInitialize{Global Flags and File System Structs}\;
\KwInitialize{Hardware Connection to Disk}

\vspace{0.5em}
\KwLoad{Superblock from \texttt{SUPERBLOCK\_OFFSET} into \texttt{Superblock\_Struct}}\;

\vspace{0.5em}
\If{EXT4 Magic Number Exists in Superblock}{
    \KwInitialize{Array of GDT Structures} 
    \KwLoad{GDT from \texttt{GDT\_OFFSET} into GDT Array}\;
    \For{Each GDT Entry}{
        \KwLoad{Inode Table from GDT Entry into Inode Array}\;      
        \For{Each Inode in Inode Array}{
            \KwLoad{Data Block Addresses from Inode}\;
            Traverse and save extents within extent trees\;
        }
    }
    \KwInitialize{Logging Framework}
}

\vspace{0.5em}
\KwRun{NBD Process}\;
\While{True}{ 
    \If{Read Operation}{
        \texttt{determine\_fs\_feature(offset)}\;
        \KwLog{file system feature accessed }\;
        \KwLoad{Data from Disk and Return to Host}\;
    }
    \If{Write Operation}{
        \texttt{determine\_fs\_feature(offset)}\;
        \KwLog{file system feature accessed}\;
        \KwLog{Changes written to feature }\;
        \KwUpdate{Local file system Structs}\;
        \KwWrite{Commit Data to Disk}\;
    }
}

\end{algorithm}

\section{Hyperparameter Sweep} \label{sec:hyperparams}

\begin{table}[ht]
\centering
\caption{Hyperparameter Search Grid} \vspace{-1em}
\label{tab:hyperparam_grids}
\resizebox{0.55\linewidth}{!}{
\setlength\extrarowheight{-3pt}
\begin{tabular}{lll}
\toprule
\textbf{Classifier} & \textbf{Parameter} & \textbf{Values} \\
\midrule
& n\_neighbors & \{1, 3, 5, 7, 10, 15, 25\} \\
\textbf{K Nearest Neighbor}& weights      & \{uniform, distance\} \\
& metric       & \{euclidean, manhattan, minkowski\} \\
\midrule
& n\_estimators & \{20, 50, 100, 200, 300\} \\
\textbf{LightGBM}& learning\_rate & \{0.01, 0.1, 1.5, 0.2, 0.3, 0.5\} \\
& max\_depth     & \{-1, 3, 5, 8, 12\} \\
\midrule
& n\_estimators & \{20, 50, 100, 200, 300\} \\
\textbf{Random Forest}& max\_depth    & \{None, 10, 20, 30, 40\} \\
& bootstrap     & \{True, False\} \\
\midrule
& C       & \{0.1, 1, 10, 100\} \\
\textbf{SVM} & gamma   & \{scale, auto\} \\
& kernel  & \{rbf, linear\} \\
\midrule
\textbf{AdaBoost}
& n\_estimators & \{20, 50, 100, 150, 200\} \\
& learning\_rate & \{0.1, 0.3, 0.5, 0.7, 1.0, 1.2\} \\
\midrule
\textbf{Naive Bayes}
& var\_smoothing & \{1e-9, 1e-8, 1e-7\} \\
\bottomrule
\end{tabular}}
\end{table}

\newpage
\section{Model Performance Table} \label{sec:perf_table_full}

\begin{figure*}[h]
\refstepcounter{figure}
\centering
\begin{tabular}{c c}
  \includegraphics[width=1.2\textwidth, angle=90, origin=c]{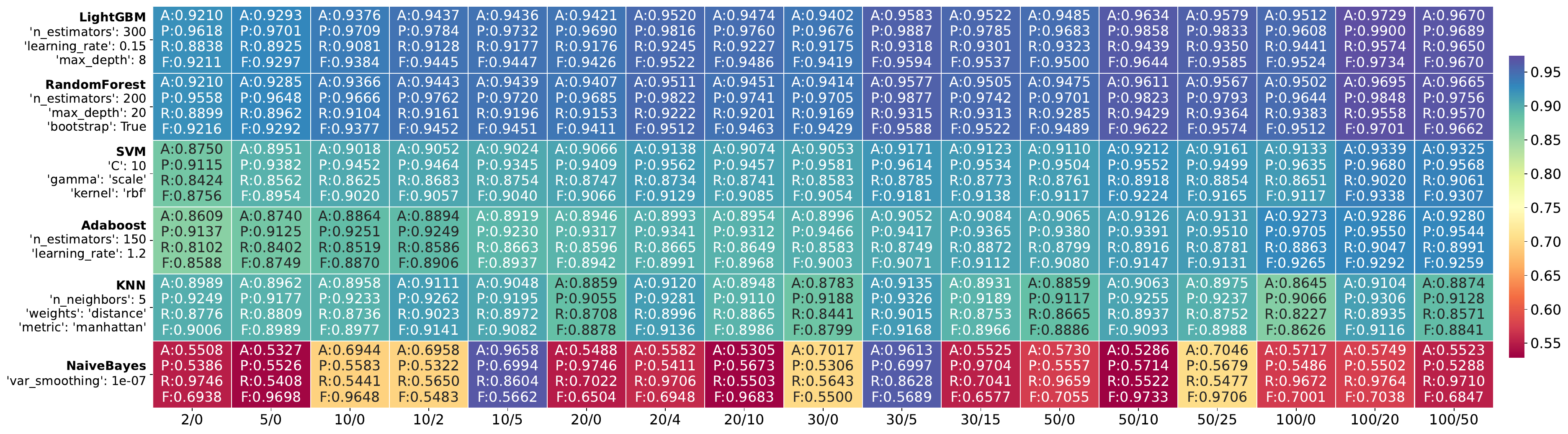}
  \includegraphics[width=1.2\textwidth, angle=90, origin=c]{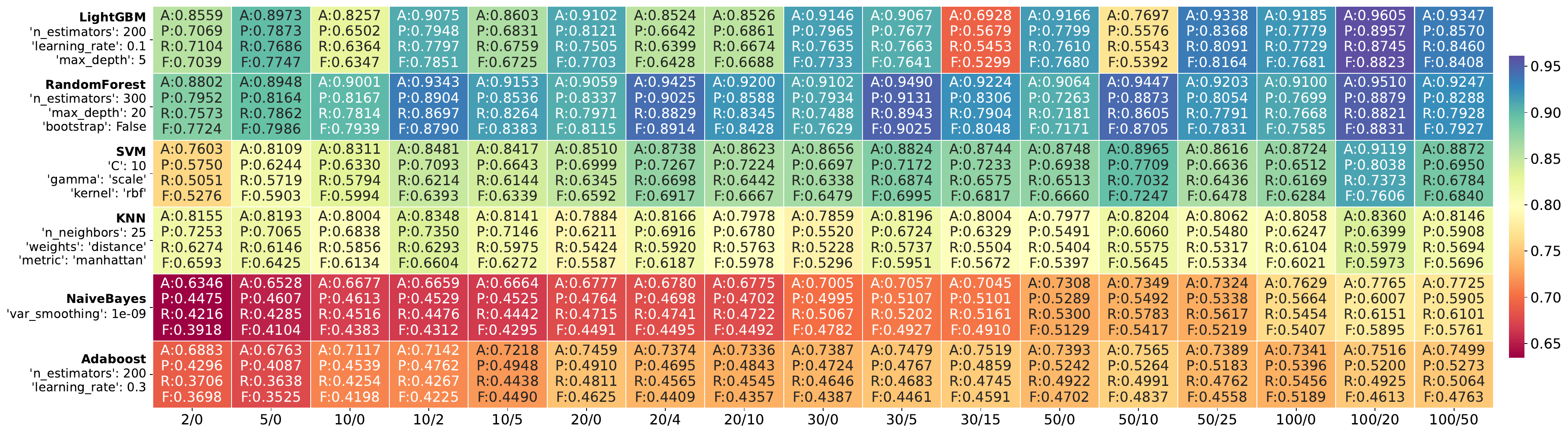}
  &
  \rotatebox[origin=c]{90}{%
    \hspace{8em}\parbox{0.70\textheight}{%
      \raggedright
      \textbf{Figure~\thefigure:} Accuracy-based gradient heatmaps for (a) binary (top) and
      (b) multiclass (bottom) classifiers, showing accuracy (A), precision (P), recall (R),
      and F1 (F) across window/overlap settings (e.g., \texttt{20/10} = 20 actions with
      10-action overlap).%
    }%
  }%
  \label{fig:heatmaps}
\end{tabular}
\end{figure*}

\end{document}